\documentclass[lettersize,journal]{IEEEtran}
\usepackage{amsmath,amssymb,amsfonts}
\usepackage{graphicx}
\usepackage{textcomp}
\usepackage{xcolor}
\usepackage{fancyhdr}
\usepackage[hyphens]{url}

\usepackage{graphicx}
\usepackage{amsmath}
\usepackage{amssymb}
\usepackage{epstopdf}
\usepackage{multirow}
\usepackage{algorithm}
\usepackage{algpseudocode}
\usepackage{hhline}
\usepackage{enumitem}
\usepackage{listings}
\usepackage{anyfontsize}
\usepackage{wrapfig}
\usepackage{footnote}
\usepackage{tablefootnote}
\usepackage{wrapfig}
\usepackage{soul}
\usepackage{tikz}
\newcommand*\circled[1]{\tikz[baseline=(char.base)]{%
            \node[shape=circle,draw,inner sep=0.5pt] (char) {#1};}}
\makesavenoteenv{tabular}
\makesavenoteenv{table}
\graphicspath{ {figures/} }
\setul{0.2ex}{1pt}

\usepackage{array}
\newcolumntype{L}[1]{>{\raggedright\arraybackslash}p{#1}}
\newcolumntype{C}[1]{>{\centering\arraybackslash}p{#1}}
\newcolumntype{R}[1]{>{\raggedleft\arraybackslash}p{#1}}

\newcolumntype{?}{!{\vrule width 1pt}}

\usepackage{xcolor,colortbl}

\newcolumntype{y}{>{\columncolor{yellow}}c}

\usepackage{fancyhdr}
\usepackage[normalem]{ulem}

\usepackage{boldline}

\definecolor{brickred}{rgb}{0.8, 0.25, 0.33}
\definecolor{dfiblue}{HTML}{4472c4}
\definecolor{dfigreen}{HTML}{70ad47}

\definecolor{comment}{HTML}{3F7F7F}
\usepackage[hidelinks]{hyperref}
\newcommand{\mydef}{\textsc{Prefender}}
\newcommand{\llydef}{Access Tracker}

\ifCLASSOPTIONcompsoc
  \usepackage[nocompress]{cite}
\else
  \usepackage{cite}
\fi
\ifCLASSINFOpdf
\else
\fi
\begin{document}
%
\title{\mydef{}: A \ul{Pref}etching Def\ul{en}der against Cache Side Channel Attacks as A Preten\ul{der}}
%
%
%
%

\author{Luyi Li,
        Jiayi~Huang,~\IEEEmembership{Member,~IEEE},
        Lang~Feng$^*$\thanks{
        Luyi Li, Lang Feng and Zhongfeng Wang are with the School of Electronic Science and Engineering, Nanjing University, Nanjing, Jiangsu 210023, China. E-mail: luyli@smail.nju.edu.cn, \{flang, zfwang\}@nju.edu.cn. Jiayi Huang is with the Department of Electrical and Computer Engineering, UC Santa Barbara, Santa Barbara, California 93106-9010 USA. E-mail: jyhuang@ucsb.edu.
        
        $^*$The corresponding authors. $^\dagger$This work was partially supported by National Natural Science Foundation of China (Grant No. 62204111) and Shuangchuang Program of Jiangsu Province (Grant No. JSSCBS20210003).},~\IEEEmembership{Member,~IEEE},
        Zhongfeng~Wang$^*$,~\IEEEmembership{Fellow,~IEEE}
        
}

\IEEEtitleabstractindextext{%

\begin{abstract}
Cache side channel attacks are increasingly alarming in modern processors due to the recent emergence of Spectre and Meltdown attacks. A typical attack performs intentional cache access and manipulates cache states to leak secrets by observing the victim's cache access patterns. Different countermeasures have been proposed to defend against both general and transient execution based attacks. Despite their effectiveness, 
they mostly trade some level of performance for security, or have restricted security scope. 
In this paper, we seek an approach to enforcing security while maintaining performance. We leverage the insight that attackers need to access cache in order to manipulate and observe cache state changes for information leakage. Specifically, we propose \mydef{}, a secure prefetcher that learns and predicts 
attack-related accesses for prefetching the cachelines to simultaneously help security and performance.
Our results show that \mydef{} is effective against several cache side channel attacks while maintaining or even improving performance for SPEC CPU 2006 and 2017 benchmarks.
\end{abstract}

}

\maketitle
\newcommand\blfootnote[1]{%
\begingroup
\renewcommand\thefootnote{}\footnote{#1}%
\addtocounter{footnote}{-1}%
\endgroup
}

\IEEEdisplaynontitleabstractindextext

%
\IEEEpeerreviewmaketitle


%
%
%
%

\section{Introduction}
\label{sec:intro}


Over the last few decades, continuing optimization of microarchitecture has led to a dramatic increase in its complexity, which might unfortunately be accompanied by many potential security vulnerabilities.
As a result, the cache side channel attacks~\cite{page2002theoretical, yarom2014flush} become serious threats to modern processors.
For example, it is possible for Spectre~\cite{kocher2019spectre} and Meltdown~\cite{lipp2018meltdown} attacks to steal almost any data in the memory, by leveraging vulnerabilities of the out-of-order execution and the speculative execution.
More seriously, these two attacks can threaten most of the modern commercial processors from Intel, AMD, and ARM.
Lots of variants of cache side channel attacks have also been found in recent years~\cite{canella2019systematic}, so the defense methods are urgently needed to enforce the security of the processors.

Cache side channel attacks exploit the cache state changes for information leakage~\cite{osvik2006cache}.
For example, the attacker can infer the cache footprint of the victim program by the time differences between cache hits and cache misses when accessing the data~\cite{kocher2019spectre, lipp2018meltdown}.
Different countermeasures have been proposed for either general or transient execution based attacks through isolation~\cite{kiriansky2018dawg}, conditional speculation~\cite{li2019conditional}, stateless mis-speculative cache accesses~\cite{yan2018invisispec}, noise injection~\cite{fang2021defeating,fang2020reuse}, prefetching{~\cite{fuchs2015disruptive},~\cite{Panda19}}, etc. However, these countermeasures either incur performance overhead, or have limited scope of security, such as only defending against the attacks conducted cross-core, so they failed to benefit both security and performance.

In this paper, we propose an approach to defeating the cache side channel attacks while maintaining or even improving the performance.
During the attack, the attacker obtains the cache state changes made by the victim by accessing the cache.
If the access patterns of both the attacker and the victim can be learned, the processor can prefetch the data that can further change the cache state to confuse the attacker.
Besides, effective prefetching can help performance if the prefetcher is able to predict the access patterns of the benign programs.


We propose \mydef{}, a prefetching defender to defeat cache side channel attacks while preserving performance benefits for benign programs. 
Specifically, three low-cost designs are proposed, which are called Scale Tracker (ST), Access Tracker (AT), and Record Protector (RP). 
Scale Tracker is able to prefetch the data that the victim may access, by tracking the target address calculation history of the memory instructions. 
Access Tracker can learn the cache access patterns of the attackers and prefetch data for confusion, even if the attackers perform intentional random accesses. 
Record Protector can link Scale Tracker and Access Tracker to prevent noisy instructions and accesses from affecting \mydef{}, and further enhance the robustness of \mydef{}.
Furthermore, effective prefetching of \mydef{} also maintains or improves performance.
The contributions of this work are as follows:
\begin{itemize}[leftmargin=2ex]
 \item \mydef{} is proposed, where a novel address prediction and a noise preventing approaches for prefetching are proposed. \mydef{} can prevent wide range of general access-based cache timing side channel attacks including both single-core and cross-core attacks, while maintaining the performance.
 \item A new approach to analyzing cache access patterns is proposed. Scale Tacker and Access Tacker are designed to realize the runtime analysis for effective prefetching.
 \item An approach is proposed to protect \mydef{} from being affected by the noisy memory instructions and accesses. To realize this, Record Protector is designed to link the scale tracker and the access tracker to help identify the cache accesses from the attackers.
 \item The detailed experiments show the effectiveness and the robustness for defeating cache side channel attacks. Besides, \mydef{} also brings the performance improvement, and is highly compatible with other prefetchers. 
\end{itemize}

For the following sections, Section~\ref{sec:background} introduces the background and the threat model. The related work is discussed in Section~\ref{sec:rework}, and the details of \mydef{} are proposed in Section~\ref{sec:design}. Then the experiments are described in Section~\ref{sec:evaluation}. Finally, Section~\ref{sec:conclusion} concludes the paper.

\section{Background and Threat Model}
\label{sec:background}

 \subsection{Cache Side Channel Attacks}
\label{subsec:side_channel}
Cache side channel attacks are to detect the cache state changes caused by the victim’s memory accesses and further infer the sensitive information of the victim from these changes. In a cache side channel attack, a round of attack is typically made up of three phases. During the first phase, the attacker initializes the cache states. For example, the attacker usually uses flush instructions to invalidate the cachelines or loads irrelative data to evict the original cachelines. Then, in the second phase, the attacker does nothing but wait for the victim to be executed. During the execution, the victim accesses its data and causes changes in the cache. In the last phase, the attacker measures which cache state is different from the initialized state and therefore deduces what data the victim has accessed.

\begin{figure}[!hbt]
 \centering
 \includegraphics[width=0.98\columnwidth]{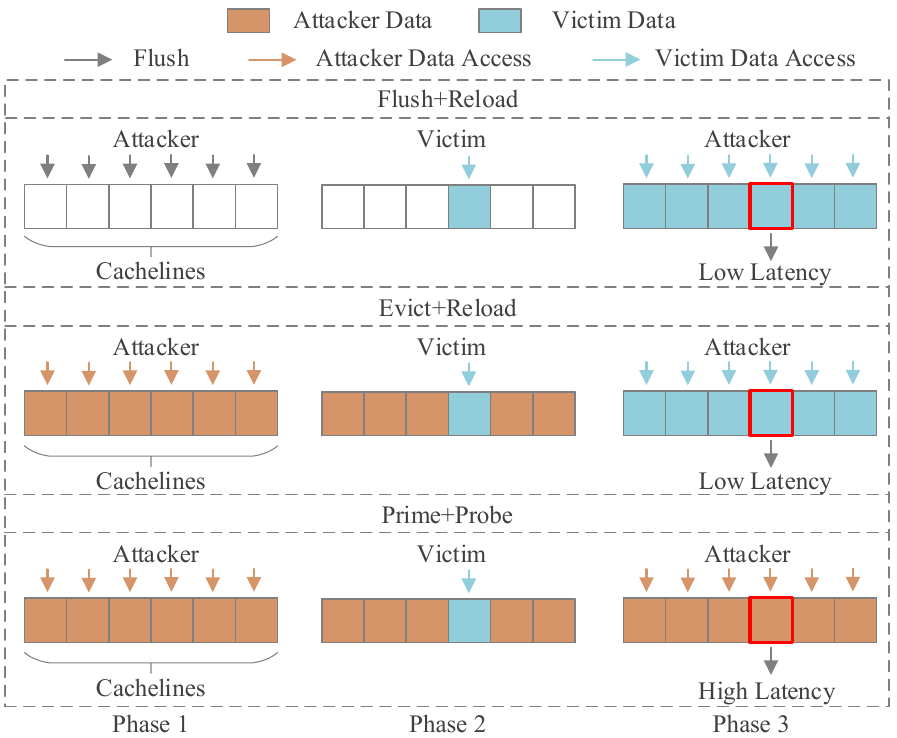}
 \caption{The examples of Flush+Reload, Evict+Reload, and Prime+Probe. The secret can be revealed by the only low (or high) latency eviction cacheline.}
 \label{fig:csca}
\end{figure}

One kind of widely used cache side channel attacks is the timing-based attack, where the cache states (hit or miss) can be identified by access latencies. Figure~\ref{fig:csca} illustrates three attacks, including Flush+Reload~\cite{yarom2014flush}, Evict+Reload~\cite{gruss2015cache}, and Prime+Probe~\cite{osvik2006cache}. Take Flush+Reload as an example, which is based on page sharing between the attacker and the victim. In phase 1, the attacker flushes all the cachelines that may be accessed by the victim. Each cacheline is called an \textit{eviction cacheline}, and they compose an \textit{eviction set}. In phase 2, the victim loads the data that are related to the secrets, which is also called secret-dependent data. In phase 3, the attacker accesses the eviction set and measures the access latency of each eviction cacheline. If the attacker detects a low latency, i.e., a cache hit, the secret might be inferred from the address of this cacheline.
For example, assuming the cacheline size is 64 bytes, if the victim loads a secret-dependent data \texttt{\small array[s$\times$64]} in phase 2, where \texttt{\small s} is the secret. During phase 3, \texttt{\small array[768]} will be accessed with a cache hit; the attacker can infer the secret is \texttt{\small s=768/64=12}.

Compared with Flush+Reload, Evict+Reload mainly differs in the way of phase 1. In Evict+Reload, the attacker loads some irrelative data to evict the cachelines instead of flush instructions. In contrast, in Prime+Probe, the attacker and the victim do not share memory pages. Therefore, the attacker has its own data which maps to the same cache sets with the victim’s data. In phase 1, the attacker evicts the cachelines by loading its own data. In phase 2, the victim accesses its data and evicts the attacker’s data. In phase 3, the attacker re-accesses its data and detects if there is a high latency, i.e., a cache miss. This cache miss can reveal the victim’s secret. 
The three attacks share the same key idea, which is to leverage the access latency to identify the secrets.

\subsection{Prefetching}
\label{subsec:prefetch}
It is widely known that the memory wall is one of the major bottlenecks of modern processors. One approach to reducing the memory access latency is prefetching, which refers to predictively loading data into the cache in advance. If the processor requests the data later, it will encounter a cache hit and the access latency is reduced. This  technique is usually implemented by the hardware module named prefetcher. Such typical examples include Tagged Prefetcher~\cite{Smith78}, Stride Prefetcher~\cite{baer1991effective}, Feedback Directed Prefetcher~\cite{srinath2007feedback}, Address Correlation Based Prefetcher~\cite{joseph1997prefetching, lai2001dead}, etc. 


\subsection{Threat Model}
\label{subsec:threat_model}
We refer to work{~\cite{He17}} to categorize the attacks. The cache timing side channel attacks that are access-based (types 2 and 4{~\cite{He17}}) are included in our threat model, which contains all the attacks described in Section{~\ref{subsec:side_channel}}.
Besides, both single-core and cross-core attacks are included.
In these attacks, the attacker is able to modify the states of any cachelines (usually the eviction cachelines) and measure their access latencies. 
The data at the eviction cachelines are either shared or conflict between the victim and the attacker.
Besides, the attacker needs and is able to access multiple eviction cachelines and leverages the timing difference between their access latencies to infer the secret of the victim\footnote{As each cacheline is not necessarily accessed multiple times in our threat model, the attack in works Prefetch-guard{~\cite{fang2021defeating}}, PrODACT{~\cite{Fang19}}, and Reuse-trap{~\cite{fang2020reuse}} is out of the scope.}.

\section{Related Work}
\label{sec:rework}
\subsection{Cache Side Channel Attacks}

\begin{table*}[!htb]
\centering
\caption{Comparisons with related work in threat model, approach, and performance overhead.}
\label{tab:compdiff}

\resizebox{0.975\textwidth}{!}{
\begin{tabular}{|c|c|c|c|c|c|c|c|c|c|c|c|c|c|c|}
\hline
 & Conditional & \multirow{1}{*}{NDA} & \multirow{1}{*}{SpecShield} & \multirow{1}{*}{InvisiSpec} & \multirow{1}{*}{SafeSpec} & \multirow{1}{*}{MuonTrap} & \multirow{1}{*}{SpecPref} & \multirow{1}{*}{Catalyst} & \multirow{1}{*}{StealthMem} & \multirow{1}{*}{DAWG} & \multirow{1}{*}{CEASER} & \multirow{1}{*}{RPcache} & \multirow{1}{*}{SHARP} & \multirow{2}{*}{\mydef{}} \\
  & Speculation~\cite{li2019conditional} & \cite{weisse2019nda} & \cite{barber2019specshield} & \cite{yan2018invisispec} & \cite{khasawneh2019safespec} & \cite{ainsworth2020muontrap} & \cite{Solanki22} & \cite{liu2016catalyst} & \cite{kim2012stealthmem} & \cite{kiriansky2018dawg} & \cite{qureshi2018ceaser} & \cite{wang2007new} & \cite{yan2017secure} &  \\\hline

 Threat Models & \multicolumn{7}{c|}{Speculative Execution Attacks} & \multicolumn{7}{c|}{ Cache Timing Side Channel Attacks} \\\hline
 Approaches & \multicolumn{3}{c|}{Speculation Restriction} & \multicolumn{4}{c|}{Shadow Structures to Hold Speculative Data} & \multicolumn{3}{c|}{Cache Partition} & \multicolumn{3}{c|}{New Cache Replacement Policy} & Prefetch \\\hline
 Performance & \multirow{2}{*}{13\%-54\%} & \multirow{2}{*}{11\%-125\%} & \multirow{2}{*}{10\%-73\%} & \multirow{2}{*}{21\%-72\%} & \multirow{2}{*}{-3\%} & \multirow{2}{*}{4\%} & \multirow{2}{*}{1.17\%} & \multirow{2}{*}{0.70\%} & \multirow{2}{*}{5.90\%} & \multirow{2}{*}{15\%} & \multirow{2}{*}{1\%} & \multirow{2}{*}{0.30\%} & \multirow{2}{*}{~0\%} & -1.69\%  \\
 Overhead &  &  &  &  &  &  &  &  &  &  &  &  &  & /-6.28\% \\\hline
\end{tabular}
}
\vspace{-2ex}
\end{table*}
 
Cache side channel attack is one of the most powerful micro-architectural side channel attacks, where the attacker can directly detect the cache states and obtain accurate timing information for inferring the secrets. Many researchers have studied various types of effective attack methods.

Kosher~\cite{kocher1996timing} firstly mentioned that the timing difference in the cache can be exploited to extract cryptographic secrets. Osvik~\cite{osvik2006cache} proposed Evict+Time and Prime+Probe methods to attack the AES algorithm~\cite{daemen1999aes}. 
In 2014, Yarom~\cite{yarom2014flush} proposed a more powerful and more fine-grained method, called Flush+Reload. This method utilizes the flush instruction supported by some architectures, for example, \texttt{clflush} in x86. Moreover, since it is based on shared memory, it has much lower noise and finer granularity, e.g., a single cacheline. The Flush+Reload has more variants, one of which is Evict+Reload~\cite{gruss2015cache}. The Evict+Reload is applicable to devices that do not support a flush instruction because it replaces the flush behavior with the cache eviction.

Research on cache side channel attacks continues to spring up, especially after Spectre~\cite{kocher2019spectre} and Meltdown~\cite{lipp2018meltdown} attacks were reported. These attacks exploit one of the most important microarchitectural optimizations, speculation, to get sensitive data. They and their variants show that many critical microarchitectural components, including Branch Target Buffer (BTB)~\cite{kocher2019spectre}, Return Stack Buffer (RSB)~\cite{Koruyeh18}, Floating Point Unit (FPU)~\cite{Stecklina18}, Page-table Entry~\cite{canella2019systematic}, Intel SGX enclave~\cite{Bulck18}, may inadvertently leak their internal states, including potential secrets while running. However, even if these attacks are based on different hardware components, most of them still leave the secrets in the cache and use cache side channel attacks such as Flush+Reload, Evict+Reload, Prime+Probe, etc., as mentioned before, to extract the information. Therefore, \mydef{} has a broad defense scale because it is able to defend against the cache side channel attacks that exploit the timing difference of cache access latency, including both traditional and transient execution based ones.

\subsection{Microarchitectural Defenses}

Many countermeasures have been proposed to defend against cache side channel attacks, including software and hardware approaches.
Software defenses are more compatible with current platforms, but they may not fundamentally defeat the attacks, and they can incur high performance overhead. Therefore, microarchitectural defenses are further proposed in many studies.

\begin{table*}[!htb]
\centering
\caption{Comparisons with related work using prefetching. (``-'' stands for that the information is not mentioned in the corresponding work.)}
\label{tab:compsimi}

\resizebox{0.98\textwidth}{!}{
\begin{tabular}{|c|c|c|c|c|c|c|c|c|}
\hline
 \multicolumn{1}{|c}{} & \multicolumn{1}{c}{} &  & \multirow{2}{*}{Prefetch-guard~\cite{fang2021defeating}} & \multirow{2}{*}{PrODACT~\cite{Fang19}} & \multirow{2}{*}{Reuse-trap~\cite{fang2020reuse}} & Disruptive & \multirow{2}{*}{BITP~\cite{Panda19}} & \multirow{2}{*}{\mydef{}} \\
\multicolumn{1}{|c}{} & \multicolumn{1}{c}{} & &  &  &  & Prefetching~\cite{fuchs2015disruptive} &  &  \\\hline
 \multirow{15}{*}{\rotatebox[origin=c]{90}{Threat Models}} & \multicolumn{8}{c|}{Access-Based Cache Attacks (Types 2 and 4~\cite{He17})} \\\cline{3-9}
  & & \multicolumn{7}{c|}{Flush+Reload} \\\cline{3-9}
  & & Single-Cacheline & $\surd$ & $\surd$ & $\surd$ & - & - & $\times$ \\\cline{3-9}
  & & Multi-Cacheline & $\times$ & $\times$ & $\times$ & - & - & $\surd$ \\\cline{3-9}
  & & \multicolumn{7}{c|}{Evict+Reload} \\\cline{3-9}
  & & Single-Cacheline & $\surd$ & - & $\surd$ & - & $\surd$ & $\times$ \\\cline{3-9}
  & & Multi-Cacheline & $\times$ & - & $\times$ & - & $\surd$ & $\surd$ \\\cline{3-9}
  & & \multicolumn{7}{c|}{Prime+Probe} \\\cline{3-9}
  & & Single-Cacheset & $\surd$ & $\surd$ & $\surd$ & - & $\surd$ & $\surd$ \\\cline{3-9}
  & & Multi-Cacheset & $\times$ & $\times$ & $\times$ & $\surd$ & $\surd$ & $\surd$ \\\cline{2-9}
  & \multicolumn{8}{c|}{Timing-Based Cache Attacks (Types 1 and 3~\cite{He17})} \\\cline{3-9}
  & & Evict+Time & \multirow{2}{*}{$\times$} & \multirow{2}{*}{$\times$} & \multirow{2}{*}{$\times$} & \multirow{2}{*}{$\times$} & \multirow{2}{*}{$\surd$} & \multirow{2}{*}{$\times$} \\
  & & Cache Collision Attack &  &  &  & &  &  \\\cline{2-9}
   & \multicolumn{8}{c|}{Single/Cross-Core Attacks} \\\cline{3-9}
  & & Single-Core & \multirow{1}{*}{$\surd$} & \multirow{1}{*}{$\surd$} & \multirow{1}{*}{$\surd$} & \multirow{1}{*}{$\surd$} & \multirow{1}{*}{$\times$} & \multirow{1}{*}{$\surd$}  \\\cline{3-9}
  & & Cross-Core & \multirow{1}{*}{$\surd$} & \multirow{1}{*}{$\surd$} & \multirow{1}{*}{$\surd$} & \multirow{1}{*}{$\surd$} & \multirow{1}{*}{$\surd$} & \multirow{1}{*}{$\surd$}  \\\hline

  
  \multicolumn{2}{|c|}{\multirow{9}{*}{\rotatebox[origin=c]{90}{Techniques}}}  & Considering & \multirow{3}{*}{$\times$} & \multirow{3}{*}{$\times$} & \multirow{3}{*}{$\surd$} & \multirow{3}{*}{$\times$} & \multirow{3}{*}{$\times$} & \multirow{3}{*}{$\surd$} \\
  \multicolumn{1}{|c}{} & & Random Access & & & & & & \\
  \multicolumn{1}{|c}{} & & Pattern & & & & & & \\\cline{3-9}
  
  \multicolumn{1}{|c}{} & & Defense & \multirow{2}{*}{Cacheline} & \multirow{2}{*}{Cacheline} & \multirow{2}{*}{Cacheline} & \multirow{2}{*}{Cacheset} & \multirow{2}{*}{Cacheline} & \multirow{2}{*}{Cacheline} \\
  \multicolumn{1}{|c}{} & &  Granularity & & & & & & \\\cline{3-9}
  
  \multicolumn{1}{|c}{} & & Handling Benign & \multirow{2}{*}{$\times$} & \multirow{2}{*}{$\times$} & \multirow{2}{*}{$\times$} & \multirow{2}{*}{$\times$} & \multirow{2}{*}{$\surd$} & \multirow{2}{*}{$\surd$} \\
  \multicolumn{1}{|c}{} & & Noise Accesses & & & & & & \\\cline{3-9}
  
  \multicolumn{1}{|c}{} & & No Software  & \multirow{2}{*}{$\times$} & \multirow{2}{*}{$\times$} & \multirow{2}{*}{$\times$} & \multirow{2}{*}{$\surd$} & \multirow{2}{*}{$\surd$} & \multirow{2}{*}{$\surd$} \\
  \multicolumn{1}{|c}{} & & Modification & & & & & & \\\hline

  \multicolumn{1}{|c}{\multirow{7}{0.1cm}{\rotatebox[origin=c]{90}{Performance}}} & \multirow{7}{*}{\rotatebox[origin=c]{90}{\& Hardware}} & \multirow{5}{2cm}{\centering Hardware Overhead} & High & High & High & Low & Low & Low  \\
  \multicolumn{1}{|c}{}& & & \multirow{1}{3cm}{\centering One conflict miss tracker and one flush instruction tracker per cache set.} & \multirow{1}{2cm}{\centering One conflict miss tracker per cache set.} & \multirow{1}{2cm}{\centering One reuse distance counter per cache set.} & \multirow{1}{3cm}{\centering One marked bit per cache set, randomization and set-balancer logic.} & \multirow{1}{2cm}{\centering BACK-INV command tracker.} & \multirow{1}{2cm}{\centering ST+AT+RP, detailed in Section~\ref{sec:hardcost}.} \\
  \multicolumn{1}{|c}{}& & & & & & & & \\
  \multicolumn{1}{|c}{}& & & & & & & & \\
  \multicolumn{1}{|c}{}& & & & & & & & \\\cline{3-9}
  \multicolumn{1}{|c}{}& & Performance & \multirow{2}{*}{-} & \multirow{2}{*}{-} & \multirow{2}{*}{-} & \multirow{2}{*}{~0\% (SPEC 2006)} & \multirow{2}{*}{1.10\% (SPEC 2006)} & \multirow{1}{*}{1.69\% (SPEC 2006)}  \\
  \multicolumn{1}{|c}{}& & Improvement & & & & & & 6.28\% (SPEC 2017) \\\hline
\end{tabular}
}
\vspace{-2ex}
\end{table*}

The comparisons between \mydef{} and related work are shown in Tables{~\ref{tab:compdiff}} and{~\ref{tab:compsimi}}.
Cache side channel attacks can be combined with transient speculative execution for data leakage, such as Spectre{~\cite{kocher2019spectre}} attacks.
To mitigate cache side channel attacks caused by transient execution, some of the prior work restricts speculation by constraining the execution of speculative loads, such as Conditional Speculation \cite{li2019conditional}, NDA~\cite{weisse2019nda} and SpecShield \cite{barber2019specshield}. They seek to identify the dangerous load instructions that can be potentially exploited by attackers and then delay their execution until all the past instructions are guaranteed to be safe. However, this method may lead to high overhead if they fail to accurately detect the dangerous \texttt{\small load}s. Another category, such as InvisiSpec~\cite{yan2018invisispec}, SafeSpec~\cite{khasawneh2019safespec} and MuonTrap~\cite{ainsworth2020muontrap}, designs a shadow structure to temporarily hold the data brought by speculative loads during transient execution, but they require many modifications to the existing hardware systems.
Although SafeSpec{~\cite{Solanki22}} achieves a 3\% performance improvement by avoiding cache pollution, its threat model is attacks on transient speculative execution, which are different from our threat model on cache timing side channel attacks.
SpecPref{~\cite{Solanki22}} also aims at speculative execution vulnerabilities and prefetchers, but the role of the prefetchers in SpecPref is the source of the data leakage instead of the way of defense, which is a different threat model.

The above defenses only prevent data leakage caused by transient execution. They are ineffective in defending against other traditional cache side channel attacks. For the traditional ones, some new cache policies were introduced. 
Catalyst~\cite{liu2016catalyst} and StealthMem~\cite{kim2012stealthmem} partition the cache into different regions for private data and shared resources, respectively. For Catalyst{~\cite{liu2016catalyst}}, software modifications are needed. DAWG~\cite{kiriansky2018dawg} achieves a higher granularity, which dynamically partitions cache ways to avoid cache sharing among different security domains. However, these methods require programmers to rewrite the source codes to flag the sensitive data. In contrast, the key idea of CEASER~\cite{qureshi2018ceaser} and RPcache~\cite{wang2007new} is to randomize the cache mapping algorithm in order to prevent the attacker from evicting the cache. 
SHARP~\cite{yan2017secure} also designs a new cache replacement policy to prevent the eviction and flush from forcing out dedicated cachelines. It requires operating system support to handle interrupts generated by alarm counters and does not defend against single-core attacks in the private cache.
Indicated in Table{~\ref{tab:compdiff}}, almost all the approaches for cache timing side channel attacks pay some level of performance for the security strength, or are not able to defeat general cache side channel attacks.
To sum up, it is always a challenging task to design both efficient and effective defenses for both security and performance. 

Besides the above studies with different approaches from \mydef{}, there are also multiple studies using prefetchers for defense, as summarized in Table{~\ref{tab:compsimi}}.
Prefetch-guard~\cite{fang2021defeating}, PrODACT~\cite{Fang19} and Reuse-trap~\cite{fang2020reuse} propose several methods to detect the spy and leverage prefetching to obfuscate the spy based on previously recorded information, sharing the same idea with \mydef{}. 
However, their threat model is different from ours. They focus on covert channel attacks. One key feature their defenses are based on is that the attacker needs to access one cacheline multiple times. As this assumption is not included in our threat model, they cannot defeat the targeting attacks of this paper.
In addition, the attacker in our threat model might access the caches randomly to mislead the prefetchers, and this is not handled by the studies{~\cite{fang2021defeating},~\cite{Fang19},~\cite{fang2020reuse}}. Moreover, no techniques are proposed in these studies to handle the noise from the benign memory accesses. These studies need software modifications and can be intrusive. For hardware consumption, since they need one tracker for each cache set, the hardware overhead can be highly increased with the growth of the cache size.
Besides, Reuse-trap needs to know the victim’s process ID in advance to record the victim’s cache misses, which may cause software modifications. 
Finally, they still trade some level of performance and fail in gaining performance improvement that can be achieved with prefetching.
Disruptive Prefetching{~\cite{fuchs2015disruptive}} also modifies the prefetchers to defeat cache side channel attacks. But it manipulates in a granularity of cacheset instead of cacheline, and only Prime+Probe is discussed, so the security is restricted.
Meanwhile, it may cause cache pollution due to its random prefetching policy.
BITP{~\cite{Panda19}} prefetches the data when identifying cross-core back-invalidation-hits in multicore systems. So, it targets cross-core attacks but not single-core attacks. In contrast, \mydef{} can also be applied to single-core attacks as it is able to filter the benign memory accesses in the single-core executions. Both BITP and \mydef{} improve performance, and \mydef{} achieves higher improvement.


Compared with related work, \mydef{} is a completely hardware-based and resource-efficient method without modifying any policy of speculative execution or cache in modern processors. It can effectively defend against the multi-cacheline (cacheset) access-based cache attacks, as well as single-core and cross-core attacks. It also considers the random accesses from the attacks and the noise from the benign accesses, and the defense granularity is each cacheline.
On the premise of ensuring security,
it further achieves a performance enhancement better than prior work through accurate runtime analyses and well-designed hardware prefetching strategies.

\section{\mydef{} Design}
\label{sec:design}

\begin{figure}[!hbt]
	\centering
	\includegraphics[width=0.98\columnwidth]{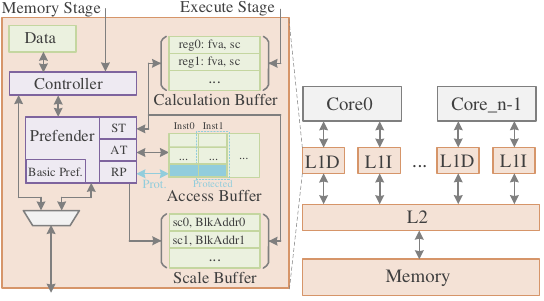}
	\caption{The overall design architecture of our system. 
	}
	\label{fig:design}
	\vspace{-2.5ex}
\end{figure}

In this section, the overview of the proposed \mydef{} shown in Figure~\ref{fig:design} is first introduced, and the details of Scale Tracker (ST), Access Tracker (AT), and Record Protector (RP) are then elaborated.

\subsection{Overview}

According to Section~\ref{subsec:side_channel}, three phases need to be performed by a cache side channel attack so the attack can be defeated by interfering with one of the phases.
\mydef{} is designed in each L1Dcache for interfering with the attacks by prefetching the eviction cachelines. Specifically, \mydef{} includes \textit{Scale Tacker (ST)} and \textit{Access Tacker (AT)} to interfere with the second and third phases, respectively. \textit{Record Protector (RP)} can further protect \mydef{} from being interfered with by the noisy memory instructions and accesses, and enhance the robustness.
A basic prefetcher (Basic Pref. in Figure~\ref{fig:design}) is also supported, such as the Tagged or Stride prefetcher. The scale tracker, the access tracker, and the basic prefetcher are able to prefetch data, while the record protector can increase the accuracy of predicting the eviction cachelines.
Note that the basic prefetcher can only help with performance, while the scale tracker, the access tracker, and the record protector can enforce security and also improve performance to some extent.

\begin{figure}[!hbt]
	\centering
	\includegraphics[width=0.98\columnwidth]{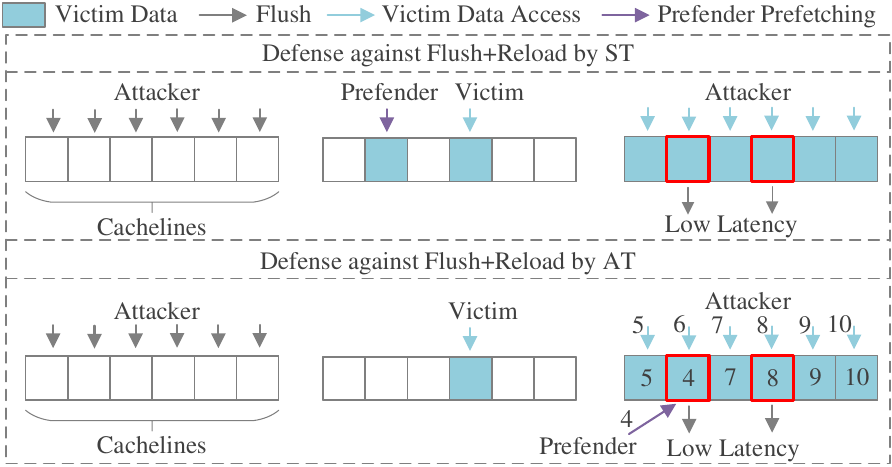}
	\caption{The example of the defenses against Flush+Reload attacks (The number near an arrow represents the access time, and the number inside each rectangle represents the first time when the corresponding cacheline is accessed).}
	\label{fig:defex}
\end{figure}

The scale tracker aims at predicting the eviction cachelines that might be accessed by the victim program during phase 2. The prediction is based on the arithmetic calculation histories of the victim instructions, which are stored in the \textit{Calculation Buffer}. The scale tracker will predict and prefetch additional eviction cachelines after a victim instruction loads the data into an eviction cacheline in phase 2. The prefetched eviction cachelines can mislead the attacker since the attacker is unable to distinguish them from the cacheline loaded by the victim instruction. An example is shown at the top of Figure~\ref{fig:defex}.

The access tracker aims at predicting the attacker's access patterns of the eviction cachelines for measuring the access latency during phase 3. The access tracker leverages the insight that a few \texttt{\small load} instructions are intensively used for the attack and stores the attacker's access patterns in the \textit{Access Buffer}. An example is shown at the bottom of Figure~\ref{fig:defex}, where the access tracker prefetches the eviction cacheline before the attacker accesses it and measures the access latency. This can also mislead the attacker.

Although the access tracker can interfere with phase 3 to mislead the attacker, since phase 3 is much longer than phase 2, there is more noise during the phase, which may affect the prediction of the access tracker.
Because phase 2 is performed by the victim, the victim's access patterns learned by the scale tracker are regarded as trusted patterns, and can help correct the prediction of the access tracker. The record protector is designed to link the scale tracker and the access tracker to prevent the noise from affecting the access tracker. The record protector can record the victim's cache access prediction of the scale tracker into the \textit{Scale Buffer}. If the attacker's access pattern in the access buffer matches a predicted victim's cache access in the scale buffer, the corresponding information in the access buffer is protected from being interfered with by the noise, and the prefetching is guided by the records in the scale buffer.

\begin{figure}[!hbt]
	\centering
	\includegraphics[width=0.98\columnwidth]{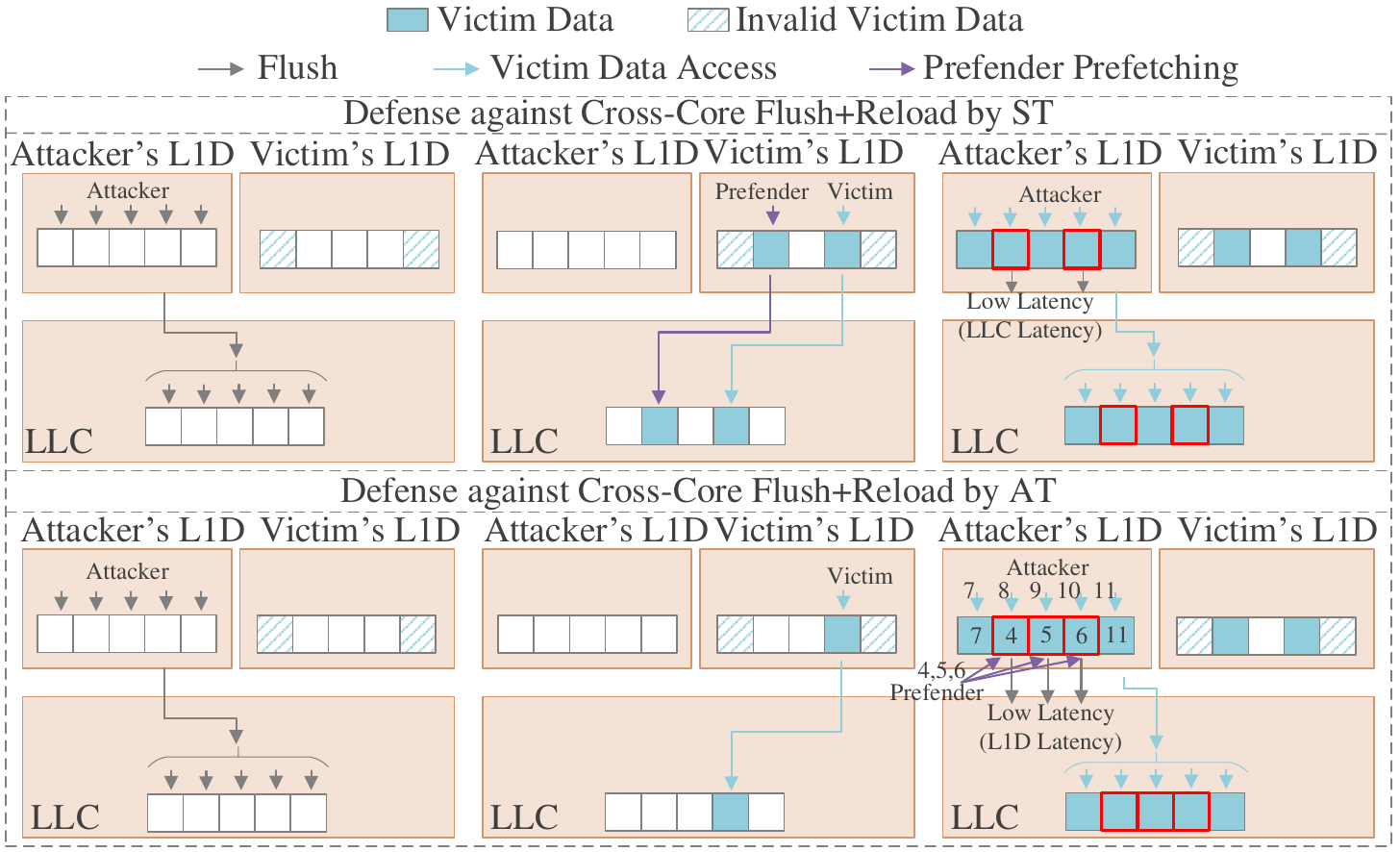}
	\caption{The example of the defenses against cross-core Flush+Reload attacks (The number near an arrow represents the access time, and the number inside each rectangle represents the first time when the corresponding cacheline is accessed).}
	\label{fig:defexcross}
\end{figure}

Note that ST and AT also work for cross-core attacks. An example is shown in Figure{~\ref{fig:defexcross}}. In this example, the programs of the attacker and the victim are on different cores with different L1D caches, but they share the same last level cache (LLC). For ST, after the attacker flush the eviction cachelines, when the victim accesses the data on another core, ST will prefetch the additional eviction cacheline similar as Figure{~\ref{fig:defex}}, both in victim's L1D cache and LLC. For phase 3, the cross-core attack originally can identify the only LLC hit to infer the sensitive information, but with ST, there are two LLC hits and the attacker is not able to distinguish which one is accessed by the victim's program. For AT, similar as the case of single-core attack, AT can directly prefetch the eviction cachelines into both attacker's L1D cache and LLC in phase 3. As the attacker keeps accessing the eviction cachelines, AT will keep prefetching, which can prefetch the cacheline in LLC accessed by the victim to L1D cache, and can directly mislead the attacker.

Since the key idea of the scale tracker, the access tracker, and the record protector is to correctly learn cache access patterns for prefetching, effective prefetching on benign loads can also improve performance while enforcing security. 
However, there are four major challenges for effective prefetching for \mydef{}.


\begin{enumerate}[itemsep=0ex,label=C\arabic*.]
    \item During phase 2, the victim may only access one eviction cacheline. Even though there are other eviction cachelines that may also be accessed, they may not be simply contiguous. How to effectively predict the access pattern given limited accesses (even single access) is challenging, which we overcome with the scale tracker.
    \item During phase 3, the eviction cachelines might be randomly accessed by the attacker. This can bypass some prefetchers such as Stride prefetcher. Predicting the eviction cachelines based on a random access pattern is challenging, which is tackled by the access tracker.
    \item During phase 3, there might be noisy memory instructions executed, so that the records of the attacker's access patterns in the access buffer are overwritten by the noisy instructions. In this case, the access tracker might be bypassed. We tackle this problem by using the record protector.
    \item During phase 3, if some non-eviction cachelines are also accessed by the same attacking instruction, the prefetching of the access tracker can be affected by these noisy accesses. We tackle this by using the record protector. Note that the challenge C3 is related to overwriting the attacker's recorded access behaviors, while the challenge C4 is about extra misleading behaviors.
\end{enumerate}

\subsection{Scale Tacker}
\label{sec:ST}



The prediction of the Scale Tracker (ST) is based on the target address calculation of the victim \texttt{\small load}. 
For example, if the \texttt{\small load}'s target address is calculated by \texttt{\small 128$\times$i+192}, where \texttt{\small i} is an integer variable, the target address can only be \texttt{\small 192}, \texttt{\small 320}, \texttt{\small 448}, etc. 
After the virtual address is translated to the physical address, if \texttt{\small paddr} is the target physical address for this time, it can be deduced that \texttt{\small paddr-128}, \texttt{\small paddr}, \texttt{\small paddr+128}, etc., may also be accessed by this \texttt{\small load} if they are in the same page.
In this way, we can predict the access pattern of the victim instruction in phase 2.
The main goal is to learn the \texttt{\small 128} as in the example, which is called the \textit{scale} in our work.

The target address of a \texttt{\small load} is usually stored in the registers, so the scale tracker needs to track how the register values are calculated.
This can be realized by recording all the calculation history of each register, but it can incur unacceptable hardware consumption. Therefore, only addition and multiplication (including subtraction and shifting) are considered, as they are widely used in the calculation, and their calculation history can be tracked by using only two values for each register.



We use two values to track the history for each register $r$: a fixed value $fva_r$ and a scale $sc_r$, which are stored in the calculation buffers. The $fva_r$ is needed to help track the scale, and it records the calculation result if all the calculations of register $r$ only depend on constant values (immediate numbers). If the value of $r$ depends on some variables such as the loaded memory values, $fva_r$ is not applicable ($NA$). 


The cache access pattern predicted by the scale tracker mainly depends on the scale $sc_r$. 
Usually, array access address in a loop is calculated as \texttt{\small base+scale$\times$i} (e.g., \texttt{\small base+128$\times$i}), where \texttt{\small base} is the base address and \texttt{\small i} is an integer variable. 
The above calculation will be propagated through some registers, and the final calculation result is stored in a register and used as the target address of a \texttt{\small load} to access the array. 
One task of the scale tracker is to track the scale by propagating scales and fixed values from registers to registers during the calculations. 
Assuming the target address $addr$ is stored in register $r$, we can obtain the scale $sc_r$ related to $r$. When one \texttt{\small load} is executed even for a single time, the scale tracker can predict that the nearby cachelines ($addr \pm sc_r$) may also be accessed by the same \texttt{\small load}. This is the access pattern tracked by the scale tracker.

The scale tracker can also support more complicated access patterns, such as \texttt{\small 128$\times$i+32$\times$j+imm}, where \texttt{\small i} and \texttt{\small j} are variables as the indices and \texttt{\small imm} is an immediate number. 
In this example, given an \texttt{\small imm}, if there is a pair of \texttt{\small i} and \texttt{\small j} makes the result to be \texttt{\small 652}, there may be another pair (e.g., \texttt{\small i} increments 1) to make the result as \texttt{\small 652+128}. 
The \texttt{\small 128} can be $sc_{r}$ in this calculation. Similarly, \texttt{\small 32} and any multiples of them like \texttt{\small 256}, \texttt{\small 512}, etc., can also be $sc_r$. 
Note that an access pattern that involves multiplications of several variables (such as \texttt{\small (128i$_0$i$_1$i$_2$+32j$_0\times$16j$_1$)$\times$(48k$_0$+imm}) ) can also be handled by propagating the scales and the fixed values during the calculations. 


\begin{table}[!htb]
\centering
\caption{\normalfont The rules to calculate $fva_{rd}$ and $sc_{rd}$. ($rd$ is the destination register; ``-''  is not applicable. $^\dagger$The rule is also for subtraction when $+$ is replaced by $-$. $^\ddagger$The rule is also for shifting when $\times$ is replaced by $>>$ or $<<$.)}
\label{tab:fvasccal}
\resizebox{0.48\textwidth}{!}{
\begin{tabular}{|c||c|c|c|c||c|c|}
\hline
 & \multicolumn{4}{c||}{Conditions} & \multicolumn{2}{c|}{Results} \\\hline
Instruction & Arg. a & Arg. b & $fva_{rs_0}$ & $fva_{rs_1}$ & $fva_{rd}$ & $sc_{rd}$ \\\hline\hline
\multirow{2}{*}{\texttt{load rd a}} & $imm_0$ & - & - & - & $imm_0$ & 1 \\\cline{2-7}
 & $imm(rs_0)$ & - & - & - & $NA$ & 1 \\\hline\hline
 \multirow{6}{*}{\texttt{add rd a b}$^\dagger$} & $rs_0$ & $imm_0$ & $NA$ & - & $NA$ & $sc_{rs_0}$  \\\cline{2-7}
 & $rs_0$ & $imm_0$ & Valid & - & $fva_{rs_0}+imm_0$ & 1  \\\cline{2-7}
 & $rs_0$ & $rs_1$ & Valid & Valid & $fva_{rs_0}+fva_{rs_1}$ & $NA$  \\\cline{2-7}
 & $rs_0$ & $rs_1$ & $NA$ & Valid & $NA$ & $sc_{rs_0}$  \\\cline{2-7}
 & $rs_0$ & $rs_1$ & Valid & $NA$ & $NA$ & $sc_{rs_1}$  \\\cline{2-7}
  & $rs_0$ & $rs_1$ & $NA$ & $NA$ & $NA$ & $min(sc_{rs_0}, sc_{rs_1})$  \\\hline\hline
\multirow{6}{*}{\texttt{mul rd a b}$^\ddagger$} & $rs_0$ & $imm_0$ & $NA$ & - & $NA$ & $sc_{rs_0}\times imm_0$  \\\cline{2-7}
 & $rs_0$ & $imm_0$ & Valid & - & $fva_{rs_0}\times imm_0$ & 1  \\\cline{2-7}
  & $rs_0$ & $rs_1$ & Valid & Valid & $fva_{rs_0}\times fva_{rs_1}$ & $NA$ \\\cline{2-7}
 & $rs_0$ & $rs_1$ & $NA$ & Valid & $NA$ & $sc_{rs_0}\times fva_{rs_1}$  \\\cline{2-7}
 & $rs_0$ & $rs_1$ & Valid & $NA$ & $NA$ & $fva_{rs_0}\times sc_{rs_1}$  \\\cline{2-7}
 & $rs_0$ & $rs_1$ & $NA$ & $NA$ & $NA$ & $sc_{rs_0}\times sc_{rs_1}$  \\\hline\hline
 Otherwise & - & - & - & - & $NA$ & 1  \\\hline

\end{tabular}
}
\end{table}

The proposed rules for calculating $sc_r$ (and $fva_r$, which can help calculate $sc_r$) are illustrated in Table~\ref{tab:fvasccal}. 
When a program is started, the fixed and scale values are initialized to $NA$ and 1, respectively.
During the execution of the program, the fixed value and scale of the destination register $rd$ are calculated according to the operand and the propagated values of the source registers.


For data movement instructions, if an immediate number is loaded to $rd$, $fva_{rd}$ is set to the number. If a value is loaded from memory to $rd$, $fva_{rd}$ and $sc_{rd}$ are reinitialized since we conservatively regard the loaded value as an unknown variable.


For addition, when $fva_{rd}$ is calculated by one immediate number and one register $rs_0$, if $rs_0$'s $fva_{rs_0}$ is $NA$, $sc_{rd}$ is the same as $sc_{rs0}$ since adding the immediate number as the offset has no effect on the scale. If $fva_{rs_0}$ is valid, $fva_{rd}$ is the addition of $fva_{rs_0}$ and the immediate number since both are fixed values.
When adding two registers, if only one of them has a valid fixed value, the scale of the destination register is the same as the scale of the source register without a valid fixed value. If neither of the source registers has a valid fixed value, the scale of the destination register can be the minimum scale of the two registers. The reason is that when the values of two registers are added, both scales can be used as the new scale. Using the minimum one can reduce the possibility of making the scale larger than a page.

For multiplication, the calculations of $fva_{rd}$ and $sc_{rd}$ are similar to those of addition, except the consideration of multiplicative factors due to multiplication. 
If any other calculations are involved, to be conservative, the destination register of the calculation is reinitialized.

When an instruction \texttt{\small load rd imm(rs)} or the equivalent instruction is executed, assuming the target address for this time is $addr'$, then $addr'\pm sc_{rs}$ are the candidate prefetching addresses. Once $sc_{rs}$ is larger than the cacheline size and smaller than the page size, the candidate addresses that are not currently in the L1Dcache are prefetched. 
We conservatively assume that all the \texttt{\small load} instructions might be the victim's instructions that are vulnerable. Therefore, the scale tracker is applied to all the \texttt{\small load} instructions.
Although all \texttt{\small load}s are considered, the defense is performed when the target addresses are calculated by addition and multiplication and the scales are larger than the cacheline size. This implies that the prefetching is performed when the \texttt{\small load}s are likely from phase 2 of the attacks instead of arbitrary \texttt{\small load}s, and this can mitigate the potential cache pollution.
For implementation, since the scale tracker prefetches data in the same page, the bitwidth for storing and calculating $fva_r$ and $sc_r$ can be small (Section~\ref{sec:hardcost}).

\begin{figure}[!hbt]
 \centering
 \includegraphics[width=0.98\columnwidth]{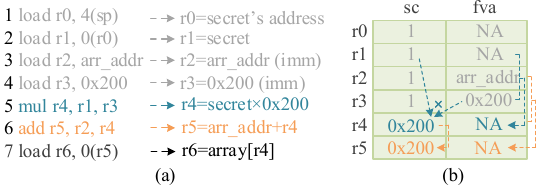}
 \caption{(a) A pseudo code example for accessing \texttt{\small array[secret$\times$0x200]}, where \texttt{\small arr\_addr} is an immediate number that represents the address of the first element in \texttt{\small array}. \\(b) The scales ($sc$) and the fixed values ($fva$) in the access buffer, where each value is set according to the instruction with the same color and the values indicated by the arrows.}
 \label{fig:STex}
\end{figure}

An example is shown in Figure~\ref{fig:STex}. The pseudo code in Figure~\ref{fig:STex}(a) accesses \texttt{\small array[secret$\times$0x200]} at line 7. For Lines 1-2, the instructions load the secret's address and the secret from the memory to r0 and r1, respectively. Therefore, the values of r0 and r1 are regarded as variables and $fva$ of them are $NA$. Lines 3-4 load the immediate numbers to r2 and r3, which makes the $fva$ of r2 and r3 be \texttt{\small arr\_addr} and \texttt{\small 0x200}, respectively. Next, line 5 multiplies r1 (\texttt{\small secret}) and r3 (\texttt{\small 0x200}) and stores the result to r4. According to Table~\ref{tab:fvasccal}, since r1's $fva$ is $NA$ and r3's $fva$ is \texttt{\small 0x200}, the $sc$ of r4 is \texttt{\small{0x200$\times$1}}, and $fva$ of r4 is $NA$. For line 6, the r2 and r4 are added to r5, which makes $sc$ of r4 directly propagated to r5 since r2 has a valid $fva$. Finally, when the \texttt{\small load} of line 7 is executed, the scale tracker will prefetch the data at \texttt{\small (target address)$\pm sc_{r5}$}, which are \texttt{\small arr\_addr+secret$\times$0x200$\pm$0x200}. In this case, assume \texttt{\small secret} is \texttt{\small 12} at this time, there are at least 2 more eviction cachelines in the cache, which can mislead the attacker to get the wrong \texttt{\small secret} value 11 or 13.

\subsection{\llydef}
\label{sec:AT}
For phase 3, the attacker needs to time all the eviction cachelines to get the access latencies. Therefore, Access  Tacker (AT) is proposed to interfere with phase 3 to further mislead the attacker. The goal of the access tracker is to learn the attacker's access pattern in phase 3, and prefetch the eviction cachelines before the attacker times them. 



However, according to challenge C2, attackers may time the eviction cachelines in a random order to bypass prefetchers such as Stride prefetcher. This increases the difficulty of learning the access patterns.
It is found that in common cases, the attacker's memory accesses in phase 3 are only associated with a few \texttt{\small load} instructions. This can help the learning of the access patterns by  recording the access history of each \texttt{\small load} instruction separately.

\begin{figure}[!hbt]
 \centering
 \includegraphics[width=0.98\columnwidth]{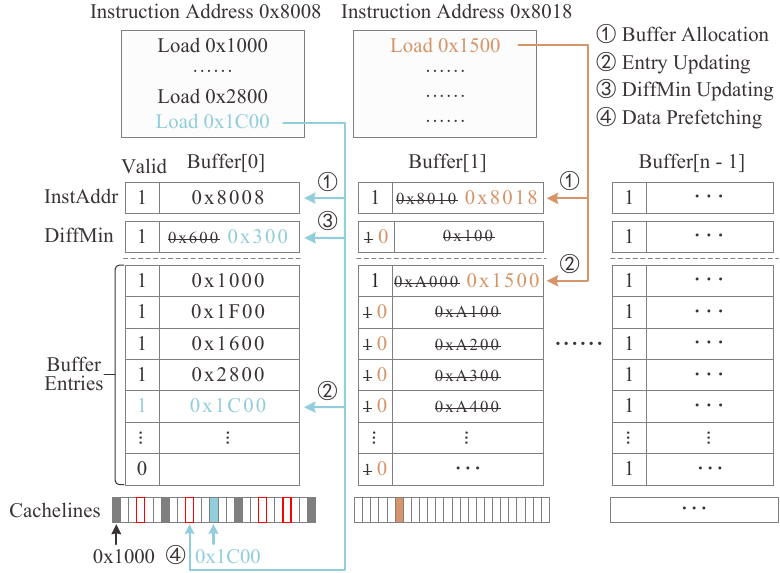}
 \caption{An example of the access buffer.}
 \label{fig:llydesign}
\end{figure}

For the access tracker, there is a set of access buffers, each of which is associated with a \texttt{\small load} instruction and records the target block addresses accessed by the associated \texttt{\small load}.
The access buffers can help the access tracker learn the access patterns of the associated \texttt{\small load} instructions. For each \texttt{\small load}, the access pattern is estimated as a stride access---an arithmetic sequence with a constant difference, which is estimated as the minimum difference between block addresses in the associated buffer.

The microarchitecture of the access buffer is shown in Figure~\ref{fig:llydesign}.
Each buffer maintains a register for storing the instruction address \texttt{\small InstAddr} of the associated \texttt{\small load}. For each entry of a buffer, the block address \texttt{\small BlkAddr} accessed by the associated \texttt{\small load} is recorded. There is also a register in each buffer, which stores the minimum difference \texttt{\small DiffMin} between two block addresses among all the entries.
Each register or entry of an access buffer has a valid bit for indicating if the data is valid or not. All valid bits are set to 0 upon the reset of the buffer.
Note that we discuss the conceptual idea in this section. For implementation, we do not need to store a complete block address in each entry (Section~\ref{sec:hardcost}).

Four stages are involved in the flow of the access tracker:


\textbf{\protect\circled{1} Buffer Allocation:} 
When a \texttt{\small load} accesses the cache each time, its instruction address (the value in the program counter) is compared with the \texttt{\small InstAddr}s to find the associated access buffer, which is then activated. If there is no associated buffer, an empty buffer is allocated to this \texttt{\small load}. If there is no empty nor associated buffer, one buffer is selected by the least recently used (LRU) replacement policy for allocation.
For example, in Figure~\ref{fig:llydesign}, when the \texttt{\small load} with \texttt{\small InstAddr 0x8008} accesses the cache, associated Buffer[0] is activated. 
In contrast, when the \texttt{\small load} with \texttt{\small InstAddr 0x8018} accesses the cache, no buffer is associated, so Buffer[1] is selected to allocate this \texttt{\small load} by LRU policy.

\textbf{\protect\circled{2} Entry Updating:} 
In the activated buffer, if the \texttt{\small BlkAddr} of the accessed data is not recorded, a new entry is selected to store this \texttt{\small BlkAddr}. If all the entries are occupied, LRU is applied to find the entry for this \texttt{\small BlkAddr}.

\textbf{\protect\circled{3} \texttt{\small DiffMin} Updating:} 
The access tracker calculates \texttt{\small DiffMin} of a buffer when the buffer is activated and the number of valid entries of this buffer surpasses a threshold (such as 4).
The number of entries of each buffer is set to be small (such as 8) to reduce the hardware complexity.
\texttt{\small DiffMin} can be used to estimate the difference between each two addresses to be accessed by the attacker in phase 3.

\textbf{\protect\circled{4} Data Prefetching:}
After the number of valid entries in a buffer surpasses a threshold (such as 4), each time this buffer is activated, candidate prefetching addresses are calculated. 
If \texttt{\small BlkAddr$'$} is the block address of the current \texttt{\small load}, the candidate prefetching addresses are \texttt{\small BlkAddr$'\pm$ DiffMin}. 
Then, the access tracker checks if these addresses exist in the activated buffer, and prefetches one of them that is not in the activated buffer nor in the L1DCache.
For example, assuming the cacheline size is 256 bytes, in Figure~\ref{fig:llydesign}, the colored cachelines' block addresses are recorded in the buffer entries, where the cachelines and their corresponding block addresses have the same color.
When \texttt{\small load} with \texttt{\small InstAddr 0x8008} accesses the cache, Buffer[0] is activated. The latest block address \texttt{\small 0x1C00} is stored in the buffer, and \texttt{\small DiffMin} is updated to \texttt{\small 0x300} as it is calculated by \texttt{\small|0x1F00-0x1C00|}.
At this moment, the access tracker predicts that the eviction cachelines are \texttt{\small 0x1C00 + 0x300$\times$k}, where \texttt{\small k} is an integer.
The red margins in Figure~\ref{fig:llydesign} indicate the eviction cachelines that are not currently accessed.
In this case, the candidate addresses are \texttt{\small 0x1C00$\pm$0x300}. As \texttt{\small 0x1C00+0x300} is already in the buffer, \texttt{\small 0x1C00-0x300} is finally prefetched by the access tracker (indicated by the arrow near \protect\circled{4}).

In this way, the access tracker can learn the access patterns of the actively executed \texttt{\small load}, and prefetch the data accordingly to mislead the attacker.
We conservatively assume that all the \texttt{\small load}s might be leveraged by the attacker, so the access tracker is applied to all of them.
The possibility of associating the buffer with the attacker's \texttt{\small load} can be increased by increasing the number of the buffers.
Note that the access tracker (or the scale tracker) only prefetch one cacheline for each \texttt{\small load} execution in order to reduce the risk of incurring performance overhead and avoid additional hardware complexity. Although all \texttt{\small load}s are considered, the access tracker finally prefetches when a \texttt{\small load} is frequently executed in a time interval, which is the access pattern of the attack's phase 3. Therefore, prefetching happens when the \texttt{\small load}s are likely from the attacks instead of arbitrary \texttt{\small load}s, and the potential cache pollution is mitigated.

\subsection{Record Protector}
The access tracker can defeat the side channel attacks by prefetching the data that are predicted to be accessed by the attackers in phase 3. However, in practice, two scenarios (challenges C3 and C4) might bypass the access tracker.
\begin{itemize}[leftmargin=2ex]
    \item Challenge C3: In phase 3, between two eviction cacheline accesses of the attacker's \texttt{load}, there might be other benign memory access instructions executed, which are noise for the access tracker. 
    The access buffer associated with the attacker's \texttt{\small load} can be occupied by a noisy instruction. According to the access tracker's policy, this noisy instruction will initialize the buffer and evict the attacker's information. In this case, the access tracker may fail to prefetch the eviction cachelines to defeat the attack.
    
    \item Challenge C4: In phase 3, if the attacker accesses the non-eviction cachelines, the access tracker will calculate wrong \texttt{\small DiffMin}. These accessed cachelines are also noise for the access tracker. For example, the \texttt{\small BlkAddr}s stored in the access buffers are \texttt{\small 0x8000}, \texttt{\small 0x8200}, \texttt{\small 0x8400}, and \texttt{\small 0x8600}, which are all the eviction cachelines. The \texttt{\small DiffMin} is \texttt{\small 0x200} in this case. However, once a non-eviction cacheline with \texttt{\small BlkAddr=0x8100} is accessed by the same \texttt{\small load}, \texttt{\small DiffMin} will be changed to \texttt{\small 0x100}. This can mislead  the access tracker to prefetch the cachelines that are not the eviction cachelines, and the attacks can bypass the access tracker's defense. 
\end{itemize}

To tackle the above two challenges, Record Protector (RP) is proposed, which can link the scale tracker and the access tracker to increase the robustness of \mydef{}, as shown in Fig~\ref{fig:design}. When a victim \texttt{\small load} accesses the cache, assuming register $r$ stores the target address, the scale tracker will prefetch the data according to $sc_r$. Meanwhile, the record protector will store $sc_r$ and the block address \texttt{\small BlkAddr$_r$} of this access's target address to the scale buffer. Each time when the attacker's \texttt{\small load} accesses the cachelines for the timing measurement in phase 3, the block address \texttt{\small BlkAddr'} of this access is checked with all the $sc_i$ and \texttt{\small BlkAddr$_i$} pairs in the scale buffer, where \texttt{\small i} is the index of the entry.
If \texttt{\small (BlkAddr'-BlkAddr$_i$)\%$sc_i$=0}, it is estimated that this access is the access to the eviction cachelines. 
Therefore, the associated access buffer is protected so that it cannot be directly replaced by LRU, and this tackles challenge C3. 
Meanwhile, upon protection, the prefetching is guided by $sc_i$ but not \texttt{\small DiffMin}, which can protect \mydef{} from being affected by the non-eviction cacheline records in the access buffer, and challenge C4 is tackled.

\begin{figure}[!hbt]
 \centering
 \includegraphics[width=0.98\columnwidth]{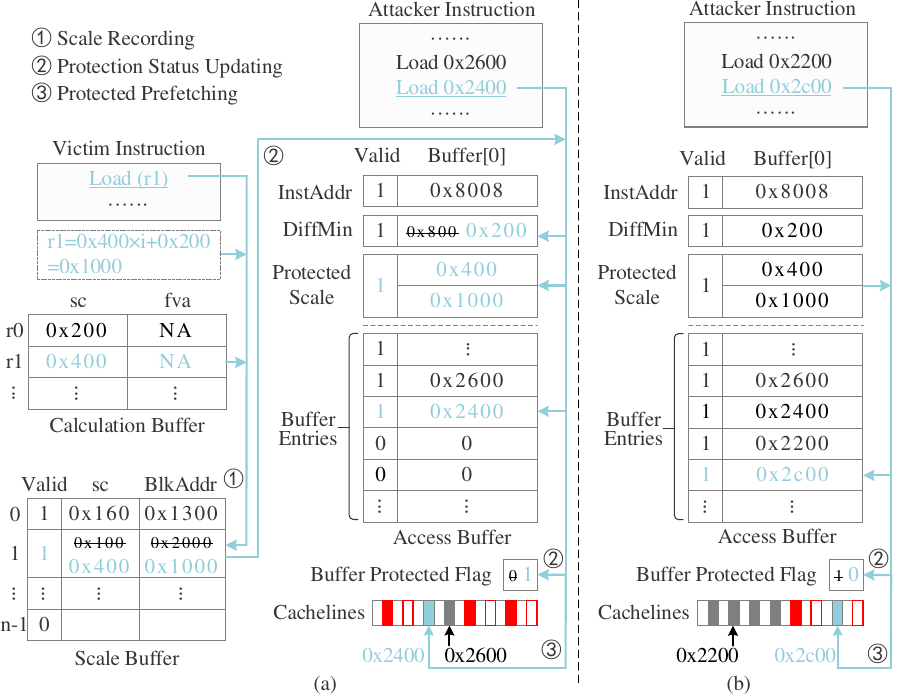}
 \caption{An example of the flow of the record protector. (The underlined instructions access the eviction cachelines; Each red margin is at the candidate address by the access tracker's policy; Each red block is at the candidate address by the record protector's policy.)}
 \label{fig:RPex}
\end{figure}

An example of the flow of the record protector is shown in Figure~\ref{fig:RPex}, and the detailed policy of the record protector is elaborated as follows, where 3 stages are involved. 

\textbf{\protect\circled{1} Scale Recording:}
When the victim accesses the eviction cacheline, the scale tracker uses the scale $sc'$ in the calculation buffer for prefetching (Section~\ref{sec:ST}). At the same time, the record protector records $sc'$ and the block address \texttt{\small BlkAddr'} of the target address to the scale buffer, as shown in step \protect\circled{1} of Figure~\ref{fig:RPex}(a).
The records in the scale buffer represent the \textit{pattern} of the possible eviction cachelines. They can guide the access tracker to avoid being affected by noisy accesses, which is discussed in the later stages.

However, one pattern might be a subset of another pattern. If so, to reduce the redundancy, only the pattern with the larger scale is recorded.
For example, in the step \protect\circled{1} of Figure~\ref{fig:RPex}(a), \texttt{\small r1} is calculated by \texttt{\small 0x400$\times$i+0x200}, and the target address is \texttt{\small 0x1000} for this time. In this case, \texttt{\small $sc'=$0x400} and \texttt{\small BlkAddr'=0x1000}, so the pattern is $S'=$\{... \texttt{\small 0x0c00}, \texttt{\small 0x1000}, \texttt{\small 0x1400}, ...\}. For Entry 1 of the scale buffer, \texttt{\small $sc_1=$0x100} and \texttt{\small BlkAddr$_1$=0x2000}, so the pattern is $S_1=$\{... \texttt{\small 0x1f00}, \texttt{\small 0x2000}, \texttt{\small 0x2100}, ...\}.
Since $S'\subset S_1$ (which means $sc'>sc_1$), all the possible eviction cachelines in $S'$ are also in $S_1$. In this case, only $S'$ needs to be kept for reducing redundancy, and Entry 1 is replaced by \texttt{\small $sc'$} and \texttt{\small BlkAddr'}.
In detail, assuming the scale and the block address related to the current \texttt{\small load} are $sc'$ and \texttt{\small BlkAddr'}, when (\texttt{\small BlkAddr'}$-$\texttt{\small BlkAddr$_i$})$\%min(sc',sc_i)=0$ for Entry i of the scale buffer, only if $sc'>sc_i$, Entry i will be updated by $sc'$ and \texttt{\small BlkAddr'}.

\textbf{\protect\circled{2} Protection Status Updating:}
In phase 3, each time when the attacker's \texttt{\small load} accesses the cache, \texttt{\small BlkAddr'} of this \texttt{\small load}'s target address is checked with all the records ($sc_i$ and \texttt{\small BlkAddr$_i$}) in the scale buffer. 
If \texttt{\small BlkAddr'} matches one of the recorded patterns, which means \texttt{\small (BlkAddr'-BlkAddr$_i$)\%$sc_i$=0}, we say \texttt{\small BlkAddr'} hits the scale buffer. 
When a cache access hits the scale buffer, it is estimated that the \texttt{\small load} of this access is the attacker's \texttt{\small load} in phase 3.
Therefore, upon the hit, the hit $sc_i$ and \texttt{\small BlkAddr$_i$} are copied to the protected scale registers in the associated access buffer, and this associated access buffer is marked as protected. 
With the record protector, the LRU policy in the access tracker for access buffer replacement is only applied to the unprotected access buffers.
By using the scale buffer to predict which \texttt{\small load} is from the attacker, and protect its associated access buffer, the access buffer will not be replaced by noisy \texttt{\small load}s. In this way, challenge C3 is tackled.

For example, in the step \protect\circled{2} of Figure~\ref{fig:RPex}(a), \texttt{\small load} accesses address \texttt{\small 0x2400}, which corresponds to an eviction cacheline. Since it hits the scale buffer, the associated access buffer is marked as protected by setting the ``Buffer Protected Flag'' as 1. Scale \texttt{\small 0x400} and block address \texttt{\small 0x1000} are also copied to the protected scale registers. 

\textbf{\protect\circled{3} Protected Prefetching:}
Besides tackling challenge C3 by protecting the access buffers, challenge C4 can be tackled by prefetching data according to the scales in the scale buffer.
Each time a \texttt{\small load}'s target block address \texttt{\small BlkAddr'} is stored into the access tracker, if it hits the scale buffer or the protected scale, the access tracker will use the hit scale $sc_{hit}$ to prefetch data, i.e., the access tracker's candidate prefetching addresses are \texttt{\small BlkAddr'$\pm sc_{hit}$}. Otherwise, the candidate prefetching addresses are calculated by the access tracker's policy in Section~\ref{sec:AT}.
So, the noisy accesses of the attacker's \texttt{\small load} have much lower effects on the defense.

An example is the step \protect\circled{3} of Figure~\ref{fig:RPex}(a). The \texttt{\small load} accesses address \texttt{\small 0x2400}. Although \texttt{\small DiffMin} in the associated buffer is \texttt{\small 0x200}, the prefetching is performed based on the hit scale \texttt{\small 0x400} in the scale buffer. As a result, one of the candidate addresses \texttt{\small 0x2400$\pm$0x400} not in the access buffer is prefetched. If the hit scale buffer entry is replaced later so that the \texttt{\small BlkAddr'} no longer hits the scale buffer, the associated access buffer's protected scale will be checked instead. If there is a hit like the case in Figure~\ref{fig:RPex}(b), the prefetching is still performed according to the hit scale \texttt{\small 0x400}. For a protected access buffer, once the number of the prefetching using the hit scale exceeds a threshold or the buffer stays untouched for a time threshold, the access buffer is set back to unprotected status, as shown in Figure~\ref{fig:RPex}(b).

In conclusion, the record protector can help the access tracker tackle challenges C3 and C4 by protecting the access buffers and performing prefetching based on the scale tracker's information, respectively. We still conservatively assume that all the \texttt{\small load}s might be the victim's and the attacker's instructions, so the record protector is applied to all of them. 
For implementation, since the access buffer stores the block addresses, the bitwidth for the modulus calculation can be small enough to be practical (Section~\ref{sec:hardcost}).

\begin{figure*}[t!]
    \begin{center}
    \includegraphics[scale=0.82]{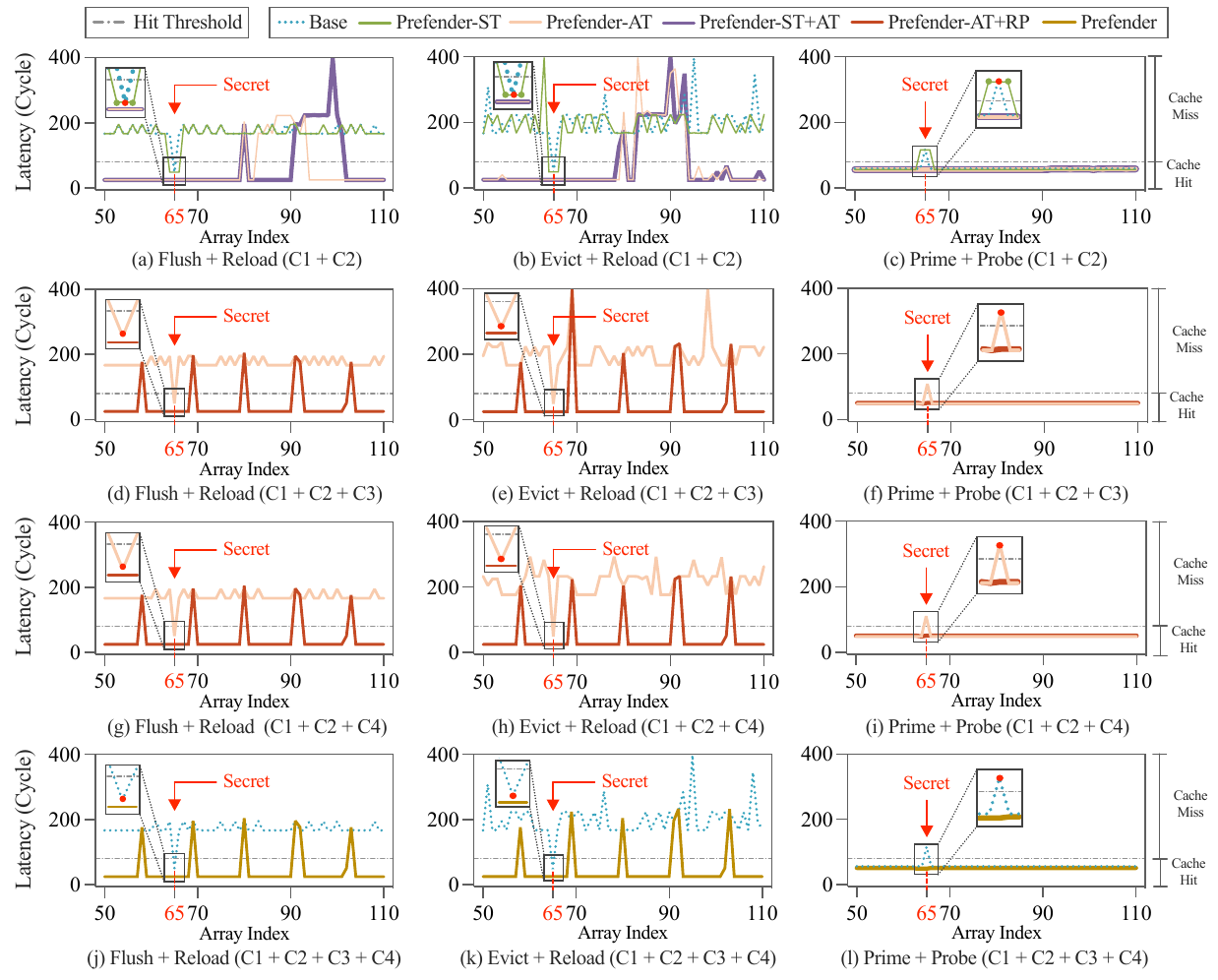}
    \end{center}
    \caption{The results of different attack methods with different challenges. (``\mydef{}'' means that the scale tracker, the access tracker, and the record protector are all applied. Note that for \mydef{}-ST, the latency results of array indices 64-66 are the same in  (a)-(c).)}
    \label{fig:security}
    \vspace{-2ex}
\end{figure*}

\begin{figure*}[h!]
    \begin{center}
    \includegraphics[scale=0.82]{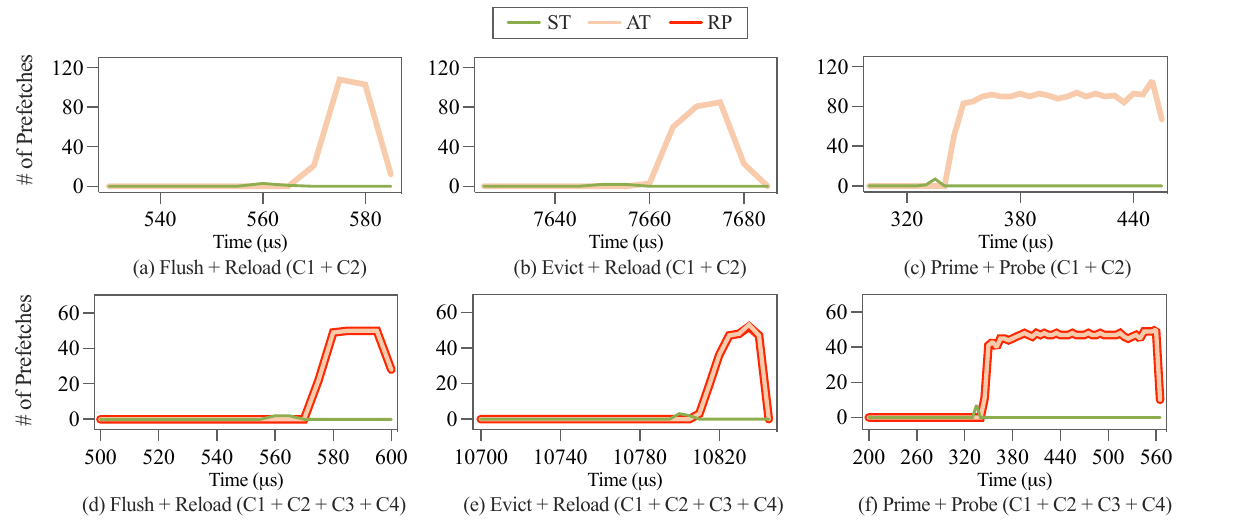}
    \end{center}
    \vspace{-2ex}
    \caption{The number of the prefetches performed under different attack methods with different challenges. (\mydef{}-ST+AT is applied in (a)-(c), and \mydef{} with the scale tracker, the access tracker, and the record protector is applied in (d)-(f). Note that the prefetches of the record protector refer to those of the access tracker guided by the record protector.)}
    \label{fig:securitydist}
\end{figure*}

\section{Evaluation}
\label{sec:evaluation}

\subsection{Experimental Setup}
In our experiments, \texttt{\small gem5} simulator~\cite{gem5} is used, where the baseline configuration contains a 2GHz x86 out-of-order CPU with a 32KB L1Icache, a 64KB L1Dcache, and a 2MB L2cache. There are 4 miss-status handling registers (MSHRs), each of which can merge at most 20 requests to the same line.
For security analysis, we test different Spectre attacks using Flush+Reload, Evict+Reload and Prime+Probe. Challenges C1-C4 are involved based on these attacks. 
For performance analysis, SPEC CPU 2006 and 2017 benchmark suites~\cite{spec,spec2017} are evaluated. 
Based upon the baseline, \mydef{} can include different basic prefetchers, including \mydef{} only, \mydef{} with a Tagged prefetcher~\cite{Smith78}, and with a Stride prefetcher~\cite{Baer91}. Note that the priority of \mydef{}'s prefetching is higher than basic prefetchers for timely defense.


\subsection{Security Evaluation}
Different side channel attacks are used to evaluate the security effectiveness of \mydef{}, and the results are shown in Figure~\ref{fig:security}.
We first evaluate without noisy memory instructions and noisy accesses (i.e., without challenges C3 and C4), and the results are shown in Figure~\ref{fig:security}(a)-(c).
For Flush+Reload, without applying \mydef{}, the attacker can infer the secret value by obtaining the only cache hit when accessing the eviction cachelines of the array in phase 3. When the Scale Tracker (ST) is applied, the scale tracker is able to introduce additional misleading cache hits on eviction cachelines, according to the calculation history. Besides, by learning the attacker's access pattern, the Access Tracker (AT) successfully predicts the accesses of the phase 3, and confuses the attacker by introducing the cache hits. When both the scale tracker and the access tracker are implemented, their effects on cachelines are overlapped. Similar results are also obtained when performing Evict+Reload attack. 
For Prime+Probe, the attacker infers the secret by the only cache miss. When the scale tracker is applied, more eviction cachelines are prefetched in phase 2, which incurs more cache misses. When the access tracker is applied, all eviction cachelines are prefetched so that the attacker can only obtain cache hits when accessing the array. This also misleads the attacker. When both the scale tracker and the access tracker are applied, only the effect of the access tracker remains since the access tracker prefetches (phase 3) after the scale tracker (phase 2).

\begin{table*}[!htb]
\centering
\caption{\normalfont Performance improvement of SPEC CPU 2006 benchmarks without the record protector. ($^\dagger$The basic prefetcher.)}
\label{tab:performance}
\resizebox{0.98\textwidth}{!}{
\begin{tabular}{|c|c||c|c|c||c|c|c|c||c|c|c|c||}
\hline
\multicolumn{2}{|c||}{Column ID} & 1 & 2 & 3 & 4 & 5 & 6 & 7 & 8 & 9 & 10 & 11 \\\hline
\multicolumn{2}{|c||}{Prefetcher} & \multicolumn{3}{c||}{\mydef{}-ST+AT} & Tagged & \multicolumn{3}{c||}{\mydef{}-ST+AT ($^\dagger$Tagged)} & Stride & \multicolumn{3}{c||}{\mydef{}-ST+AT ($^\dagger$Stride)}\\\hline
\multicolumn{2}{|c||}{\# of Acc. Tra. Buf.} & 16 & 32 & 64 & - & 16 & 32 & 64 & - & 16 & 32 & 64\\\hline

\multirow{13}{*}{\rotatebox[origin=c]{90}{Benchmark}}
 & 400.perlbench & 0.677\% & 0.679\% & 1.110\% & 0.241\% & 0.427\% & 0.588\% & 0.324\% & 0.389\% & 1.117\% & 1.065\% & 1.536\%\\\cline{2-13}
 & 401.bzip2 & 3.314\% & 3.346\% & 3.407\% & 4.428\% & 5.717\% & 5.728\% & 5.732\% & 1.777\% & 3.922\% & 3.959\% & 4.052\%\\\cline{2-13}
 & 429.mcf & 6.421\% & 8.562\% & 8.585\% & 8.636\% & 12.069\% & 12.228\% & 12.237\% & 13.233\% & 14.803\% & 17.684\% & 17.653\%\\\cline{2-13}
 & 445.gobmk & -0.025\% & -0.106\% & -0.122\% & 1.318\% & 1.164\% & 1.103\% & 1.102\% & 0.363\% & 0.433\% & 0.379\% & 0.347\%\\\cline{2-13}
 & 456.hmmer & 0.830\% & 0.862\% & 0.891\% & 10.115\% & 10.128\% & 10.152\% & 10.158\% & 7.119\% & 6.417\% & 6.474\% & 6.512\%\\\cline{2-13}
 & 458.sjeng & -0.354\% & -0.355\% & -0.366\% & -0.437\% & -0.613\% & -0.615\% & -0.609\% & -0.016\% & -0.300\% & -0.303\% & -0.322\%\\\cline{2-13}
 & 462.libquantum & 6.533\% & 6.533\% & 6.532\% & 4.852\% & 6.501\% & 6.501\% & 6.501\% & 7.555\% & 9.768\% & 9.770\% & 9.773\%\\\cline{2-13}
 & 464.h264ref & 0.269\% & 0.256\% & 0.408\% & 1.762\% & 1.707\% & 1.521\% & 1.804\% & 0.934\% & 0.724\% & 0.993\% & 0.793\%\\\cline{2-13}
 & 471.omnetpp & -0.006\% & -0.006\% & -0.011\% & 0.112\% & 0.109\% & 0.109\% & 0.109\% & 0.229\% & 0.213\% & 0.213\% & 0.211\%\\\cline{2-13}
 & 473.astar & 0.033\% & 0.398\% & -0.132\% & 0.183\% & 0.212\% & 0.415\% & -0.176\% & 0.032\% & 0.059\% & 0.474\% & -0.021\%\\\cline{2-13}
 & 483.xalancbmk & 0.702\% & 2.840\% & 3.941\% & 11.576\% & 11.577\% & 11.952\% & 10.592\% & 2.137\% & 2.771\% & 4.952\% & 5.683\%\\\cline{2-13}
 & 999.specrand & 0.000\% & 0.000\% & 0.000\% & 0.001\% & 0.001\% & 0.001\% & 0.001\% & 0.000\% & 0.000\% & 0.000\% & 0.000\%\\\cline{2-13}
 & Avg. & \textbf{1.533\%} & \textbf{1.918\%} & \textbf{2.020\%} & \textbf{3.566\%} & \textbf{4.083\%} & \textbf{4.140\%} & \textbf{3.981\%} & \textbf{2.813\%} & \textbf{3.327\%} & \textbf{3.805\%} & \textbf{3.851\%}\\\hline

\end{tabular}
}
\vspace{-2ex}
\end{table*}

\begin{table*}[!htb]
\centering
\caption{\normalfont Performance improvement of SPEC CPU 2006 benchmarks with the record protector. ($^\dagger$The basic prefetcher.)}
\label{tab:performanceRP}
\resizebox{0.98\textwidth}{!}{
\begin{tabular}{|c|c||c|c|c||c|c|c|c||c|c|c|c||}
\hline
\multicolumn{2}{|c||}{Column ID} & 1 & 2 & 3 & 4 & 5 & 6 & 7 & 8 & 9 & 10 & 11 \\\hline
\multicolumn{2}{|c||}{Prefetcher} & \multicolumn{3}{c||}{\mydef{}} & Tagged & \multicolumn{3}{c||}{\mydef{} ($^\dagger$Tagged)} & Stride & \multicolumn{3}{c||}{\mydef{} ($^\dagger$Stride)}\\\hline
\multicolumn{2}{|c||}{\# of Acc. Tra. Buf.} & 16 & 32 & 64 & - & 16 & 32 & 64 & - & 16 & 32 & 64\\\hline

\multirow{13}{*}{\rotatebox[origin=c]{90}{Benchmark}}
 & 400.perlbench & 0.584\% & 0.562\% & 0.585\% & 0.241\% & 0.001\% & 0.524\% & 0.545\% & 0.389\% & 1.115\% & 1.118\% & 1.116\%\\\cline{2-13}
 & 401.bzip2 & 3.129\% & 3.192\% & 3.251\% & 4.428\% & 5.621\% & 5.646\% & 5.667\% & 1.777\% & 3.828\% & 3.916\% & 3.958\%\\\cline{2-13}
 & 429.mcf & 4.347\% & 5.494\% & 5.497\% & 8.636\% & 9.335\% & 9.557\% & 9.540\% & 13.233\% & 12.114\% & 12.755\% & 12.755\%\\\cline{2-13}
 & 445.gobmk & -0.030\% & -0.066\% & -0.084\% & 1.318\% & 1.189\% & 1.171\% & 1.163\% & 0.363\% & 0.386\% & 0.347\% & 0.335\%\\\cline{2-13}
 & 456.hmmer & 0.830\% & 0.861\% & 0.891\% & 10.115\% & 10.128\% & 10.149\% & 10.162\% & 7.119\% & 6.431\% & 6.467\% & 6.529\%\\\cline{2-13}
 & 458.sjeng & -0.411\% & -0.428\% & -0.422\% & -0.437\% & -0.649\% & -0.660\% & -0.687\% & -0.016\% & -0.324\% & -0.337\% & -0.373\%\\\cline{2-13}
 & 462.libquantum & 6.516\% & 6.518\% & 6.521\% & 4.852\% & 6.502\% & 6.502\% & 6.502\% & 7.555\% & 9.781\% & 9.782\% & 9.782\%\\\cline{2-13}
 & 464.h264ref & 0.346\% & 0.300\% & 0.346\% & 1.762\% & 1.739\% & 1.806\% & 1.800\% & 0.934\% & 0.899\% & 0.812\% & 0.843\%\\\cline{2-13}
 & 471.omnetpp & 0.025\% & 0.047\% & 0.058\% & 0.112\% & 0.104\% & 0.112\% & 0.106\% & 0.229\% & 0.231\% & 0.225\% & 0.230\%\\\cline{2-13}
 & 473.astar & 0.029\% & 0.308\% & -0.139\% & 0.183\% & 0.208\% & 0.355\% & -0.182\% & 0.032\% & 0.054\% & 0.385\% & -0.027\%\\\cline{2-13}
 & 483.xalancbmk & 0.860\% & 2.372\% & 3.822\% & 11.576\% & 11.533\% & 11.704\% & 10.624\% & 2.137\% & 3.123\% & 4.644\% & 5.628\%\\\cline{2-13}
 & 999.specrand & 0.000\% & 0.000\% & 0.000\% & 0.001\% & 0.001\% & 0.001\% & 0.001\% & 0.000\% & 0.000\% & 0.000\% & 0.000\%\\\cline{2-13}
 & Avg. & \textbf{1.352\%} & \textbf{1.597\%} & \textbf{1.694\%} & \textbf{3.566\%} & \textbf{3.809\%} & \textbf{3.905\%} & \textbf{3.770\%} & \textbf{2.813\%} & \textbf{3.136\%} & \textbf{3.343\%} & \textbf{3.398\%}\\\hline

\end{tabular}
}
\vspace{-3ex}
\end{table*}

\begin{table*}[!htb]
\centering
\caption{\normalfont Performance improvement of SPEC CPU 2017 benchmarks. ($^\dagger$The basic prefetcher.)}
\label{tab:performance2017}
\resizebox{0.98\textwidth}{!}{
\begin{tabular}{|c|c||c|c||c|c|c||c|c|c||}
\hline
\multicolumn{2}{|c||}{Column ID} & 1 (ST+AT) & 2 & 3 & 4 (ST+AT) & 5 & 6 & 7 (ST+AT) & 8   \\\hline
\multicolumn{2}{|c||}{Prefetcher} & \multicolumn{2}{c||}{\mydef{}} & Tagged & \multicolumn{2}{c||}{\mydef{} ($^\dagger$Tagged)} & Stride & \multicolumn{2}{c||}{\mydef{} ($^\dagger$Stride)}\\\hline
\multicolumn{2}{|c||}{\# of Acc. Tra. Buf.} & 32 & 32 & - & 32 & 32 & - & 32 & 32  \\\hline

\multirow{10}{*}{\rotatebox[origin=c]{90}{Benchmark}}
 & 507.cactuBSSN\_r & 0.917\% & 0.874\% & 12.256\% & 12.752\% & 12.711\% & 10.707\% & 11.672\% & 11.557\%\\\cline{2-10}
 & 526.blender\_r & 0.015\% & 0.015\% & 0.356\% & 0.302\% & 0.302\% & 0.120\% & 0.133\% & 0.133\%\\\cline{2-10}
 & 531.deepsjeng\_r & -0.396\% & -0.379\% & -0.121\% & -0.525\% & -0.513\% & 0.000\% & -0.380\% & -0.369\%\\\cline{2-10}
 & 538.imagick\_r & 5.664\% & 5.664\% & 4.240\% & 6.389\% & 6.389\% & 0.561\% & 6.292\% & 6.292\%\\\cline{2-10}
 & 541.leela\_r & -0.072\% & -0.249\% & 0.164\% & 0.257\% & 0.120\% & 0.145\% & 0.187\% & 0.073\%\\\cline{2-10}
 & 557.xz\_r & 0.243\% & 0.332\% & 4.015\% & 4.107\% & 4.104\% & 1.637\% & 1.873\% & 1.892\%\\\cline{2-10}
 & 510.parest\_r & 39.738\% & 50.291\% & 44.043\% & 49.822\% & 54.617\% & 0.700\% & 35.586\% & 46.775\%\\\cline{2-10}
 & 548.exchange2\_r & 0.000\% & -0.006\% & 0.000\% & 0.000\% & 0.000\% & 0.011\% & -0.004\% & 0.015\%\\\cline{2-10}
 & 554.roms\_r & 0.000\% & 0.000\% & 30.898\% & 30.898\% & 30.898\% & 15.797\% & 15.797\% & 15.797\%\\\cline{2-10}
 & Avg. & \textbf{5.123\%} & \textbf{6.282\%} & \textbf{10.650\%} & \textbf{11.556\%} & \textbf{12.070\%} & \textbf{3.298\%} & \textbf{7.906\%} & \textbf{9.129\%}\\\hline

\end{tabular}
}
\vspace{-3ex}
\end{table*}

When there are noisy memory instructions during phase 3 (challenge C3), the access buffers of the access tracker can be occupied by these accesses of the noisy instructions, and applying the access tracker only may not defeat the attack, as shown in Figure~\ref{fig:security}(d)-(f). However, when the Record Protector (RP) is implemented, the access buffer associated with the attacker's \texttt{\small load} is successfully identified and protected, so the access tracker is able to prefetch the eviction cachelines and mislead the attacker again. 
Similarly, without the record protector, when there are noisy accesses by the attacker's \texttt{\small load} in phase 3 (challenge C4), the value of \texttt{\small DiffMin} can be affected, and the access tracker may not be able to prefetch the eviction cachelines, as shown in Figure~\ref{fig:security}(g)-(i). In contrast, when the record protector is applied, the prefetching is guided by the scale buffer that contains the possible eviction cachelines from the victim, so the access tracker can again correctly prefetch the eviction cachelines to mislead the attacker.

Combining all the challenges and all the designs, the security can be illustrated in Figure~\ref{fig:security}(j)-(l). Without applying \mydef{}, the attacker can infer the secret with the only one cache hit (or miss). With \mydef{}, even though all the challenges are involved, multiple cache hits (or misses) are introduced, and the attack is defeated.

We further analyze the insights of the defense, which are shown in Figure~\ref{fig:securitydist}, where the x-axis represents the execution time. We only show the part where the attack is performed.
For Figure~\ref{fig:securitydist}(a)-(c), challenges C1 and C2 are involved, and \mydef{}-ST+AT is applied. One can notice that the scale tracker prefetches a small amount of data shown in Figure~\ref{fig:securitydist}(a)-(c), which corresponds to the data at array indices 64 and 66 of the green curves in Figure~\ref{fig:security}(a)-(c). After this, the access tracker prefetches more data shown in Figure~\ref{fig:securitydist}(a)-(c), which is also shown by the orange curves in Figure~\ref{fig:security}(a)-(c).
For Figure~\ref{fig:securitydist}(d)-(f), all challenges are involved, and \mydef{} with all three designs is applied. It is indicated that the scale tracker still prefetches several data. After this, with the guidance of the record protector, the access tracker successfully prefetches the data even with the noisy instructions and accesses. The corresponding results are shown in Figure~\ref{fig:security}(j)-(l). This further shows the mechanism of the defense.


In summary, by successfully defeating the attacks in the threat model, \mydef{} can enforce the security as the same as the previous work~\cite{li2019conditional,barber2019specshield,yan2018invisispec,khasawneh2019safespec}.



\begin{figure*}[htb!]
	\centering
	\includegraphics[width=1.98\columnwidth]{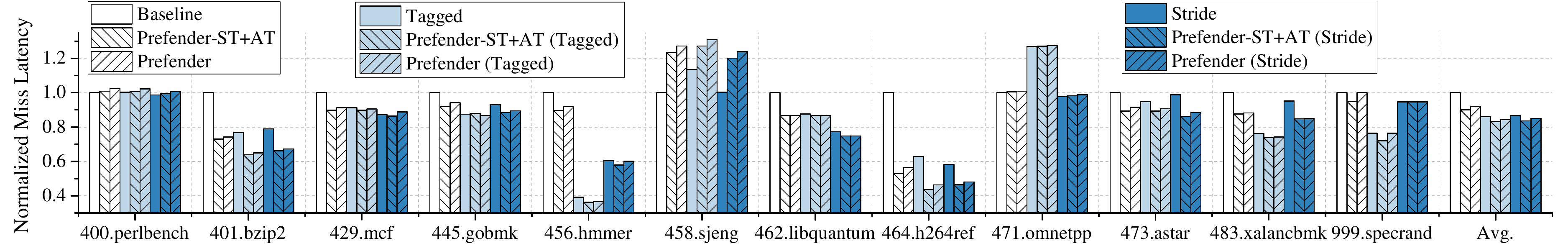}
	\caption{The normalized total latency of all cache misses of L1Dcache.}
	\label{fig:cachemisslat}
	\vspace{-3ex}
\end{figure*}

\begin{figure*}[!htb]
	\centering
	\includegraphics[width=1.98\columnwidth]{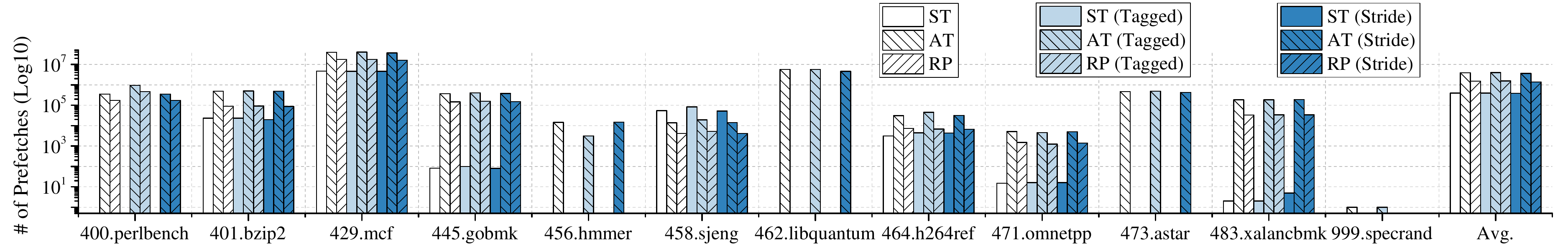}
	\vspace{-1ex}
	\caption{The number of the prefetches. (The prefetches of the record protector refer to those of the access tracker guided by the record protector.)}
	\label{fig:deftimesRP}
	\vspace{-2ex}
\end{figure*}

\subsection{Performance Evaluation}

While enforcing security, \mydef{} can also maintain or even improve performance. When the record protector is not implemented, the performance results of SPEC CPU 2006 benchmarks are shown in Table~\ref{tab:performance}. The results show the improvement percentile compared with the baseline that has no prefetchers. The main results are Columns 2, 6 and 10, where 32 access buffers are implemented. When \mydef{}-ST+AT is implemented (Column 2), the performance improvement is about 2\% on average, with the security enforcement. For Columns 6 and 10 where the conventional prefetchers are applied, \mydef{}-ST+AT can further improve the performance compared with Columns 4 and 8 where no \mydef{} is implemented, respectively. This shows \mydef{}'s capability for maintaining or even improving the performance.

When the record protector is implemented, the performance results of SPEC CPU 2006 benchmarks are shown in Table~\ref{tab:performanceRP}, where the performance distributions are similar as Table~\ref{tab:performance}. With the record protector, \mydef{} also improves the performance on average, no matter if there are basic prefetchers or not. At the same time, not only is the security enforced, but also the robustness of \mydef{} is greatly improved by the record protector.



While the performance is improved by \mydef{} on average, the impacts on different benchmarks vary.
For example, \textit{401.bzip2}, \textit{429.mcf} and \textit{462.libquantum} have the most speedup with \mydef{}. In contrast, there is almost no performance impact on \textit{999.specrand}. For \textit{445.gobmk}, \textit{458.sjeng} and \textit{471.omnetpp}, their performance has a slight drop with \mydef{}.
The effect of the number of the access buffers is also evaluated and shown in Tables~\ref{tab:performance} and \ref{tab:performanceRP}. The results indicate that more access buffers usually help the performance. Besides, if the buffers are more than 32, marginal improvements are obtained.

Besides, the results of the cases newly presented in SPEC CPU 2017 benchmarks are shown in Table{~\ref{tab:performance2017}}. Similar to SPEC CPU 2006, \mydef{} also has performance improvement, both with and without the record protector. At the same time, \mydef{} can further increase the performance based on the basic prefetchers. Note that for some benchmarks such as {\textit{510.parest\_r}}, the performance improvement is relatively large. This is because the data prefetched by \mydef{} can greatly help reduce the cache miss rate. For example, the cache miss rate and the cache misses' access latency of {\textit{510.parest\_r}} in Column 2 of Table{~\ref{tab:performance2017}} (in the revision letter) are 50.26\% and 55.99\% less than that without \mydef{}, respectively.

\subsection{Analysis of Cache Miss and Defense}
\label{sec:cmiss}

Prefetching can impact the cache miss rate and latency. We evaluated the total latency of all cache misses of each benchmark, which is shown in Figure~\ref{fig:cachemisslat}. Each result is normalized to the baseline. In Figure~\ref{fig:cachemisslat}, ``\mydef{}-ST+AT'' have the same configuration as Columns 2, 6, 10 in Table~\ref{tab:performance}, where the record protector is not applied. ``\mydef{}'' has the same configuration as Columns 2, 6, 10 in Table~\ref{tab:performanceRP} with the scale tracker, the access tracker, and the record protector. It is indicated that the total latencies of cache misses are reduced on average when \mydef{} is implemented. For a few cases, the latency becomes higher than the baseline, which leads to a slight performance drop, such as \textit{458.sjeng}. Some cases have similar miss latencies before and after applying \mydef{}, but the performance is still improved, such as \textit{400.perlbench} and \textit{429.mcf}. 



\begin{figure}[!hbt]
	\centering
	\includegraphics[width=0.98\columnwidth]{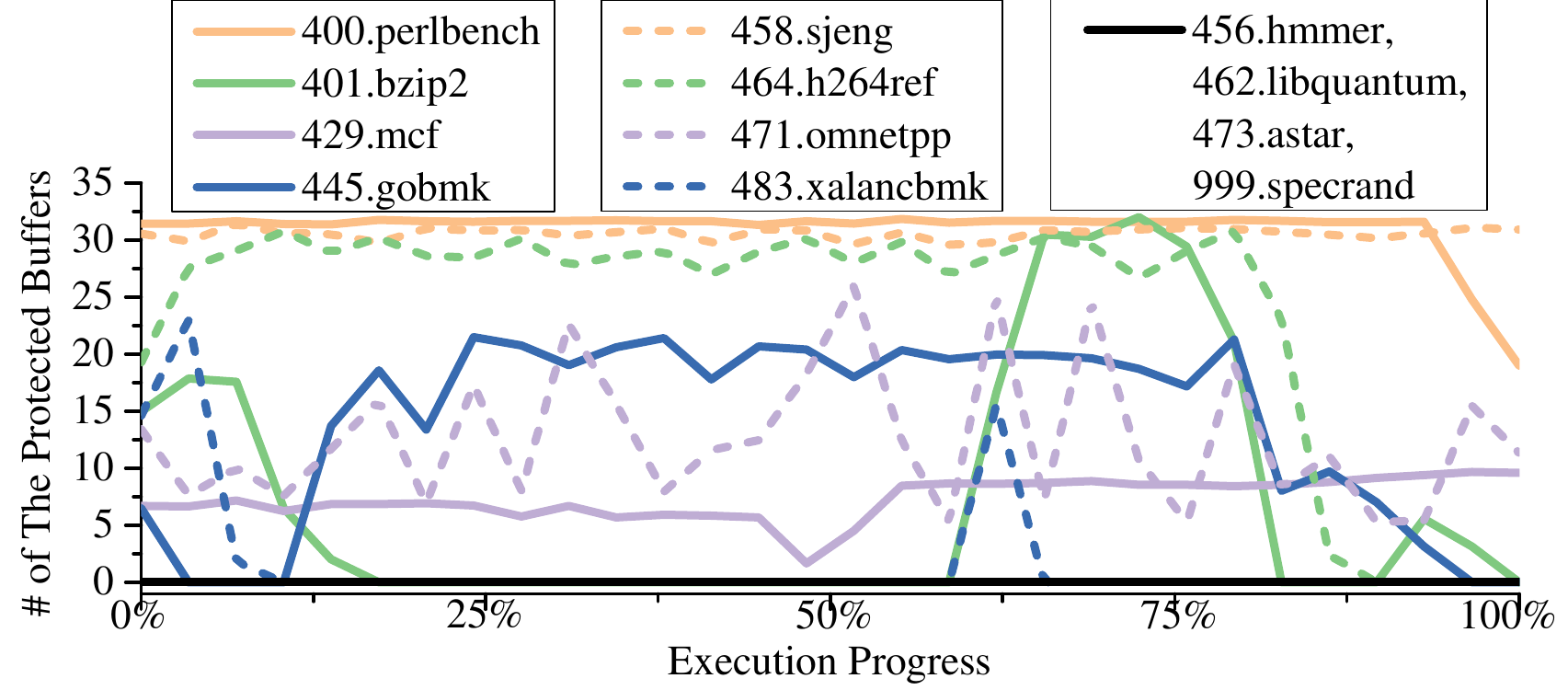}
	\vspace{-1ex}
	\caption{The number of the protected buffers during the execution. (The configurations are the same as that of Column 2 in Table~\ref{tab:performanceRP}.)}
	\label{fig:protbufnum}
\end{figure}

We further evaluated the number of the prefetches performed by the scale tracker, the access tracker, and the record protector of \mydef{}. The results are shown in Figure~\ref{fig:deftimesRP}. Note that the access tracker prefetches the most data, and the record protector guides the access tracker to prefetch more data than the scale tracker. The reason is that the scale tracker performs one prefetch when a target address of a \texttt{\small load} is calculated by addition and multiplication, and the scale is larger than the cacheline size. This happens less frequently than triggering the record protector, which helps the access tracker prefetch each time a scale from the scale tracker is recorded and a \texttt{\small load}'s target address hits the scale history. For the access tracker, the requirement for prefetching is the easiest to be satisfied since it only needs a \texttt{\small load} to be frequently executed. 

Finally, the number of the protected access buffers during the execution is tested in Figure~\ref{fig:protbufnum}, which indicates that different benchmarks have different patterns on the protected buffer numbers. For \textit{400.perlbench},  \textit{458.sjeng}, and \textit{464.h264ref}, most of the buffers are protected during the execution. In contrast, \textit{456.hmmer}, \textit{462.libquantum}, \textit{473.astar}, and \textit{999.specrand} have no protected buffer. For other benchmarks, the number of the protected buffers varies. These results indicate that different functionality of the program can affect the behaviors of the record protector.

\subsection{Hardware Resource Consumption Analysis}
\label{sec:hardcost}

We briefly analyze the upper bound of the hardware resource consumption. 
For the SRAM size of the scale tracker, the prefetching is performed within one page, so 16 bits are enough for the values in the calculation buffers even with a page size of 64KB. For each register, there are two values associated, so the scale tracker needs hundreds of bytes in total for dozens of registers. For the datapath of the scale tracker, an adder, a multiplier and a comparator are used, which are also 16-bit.

For the SRAM size of the access tracker, there are 32 access buffers, each of which has 8 entries. Even if each value of the buffer is 64-bit, only $<$3KB SRAMs are required. For the datapath of the access tracker, since the access tracker predicts and prefetches the eviction cachelines, 20 bits are enough for calculating the \texttt{\small DiffMin} even when L1Dcache is as large as 1MB. Several 20-bit comparators and 20-bit adders are used for each access buffer. The hardware consumption is also reasonable.

For the SRAM size of the record protector, the scale buffer has 8 entries in the experiments, with each entry $16$($sc$)$+64$(\texttt{\small BlkAddr})$=80$ bits. For each access buffer, the record protector requires another 80-bit register for the scale history. Therefore, 400 bytes are needed. For the datapath of the record protector, a 2-way associative L1Dcache is 64KB, with each cacheline of 64 bytes, so 9 bits are used for the set index of the cache. Since the target of the prefetching is the cachelines, we only use the set index (the address of the cache entries) to calculate the modules, and several hardware modules of 9-bit modulus are needed. According to the synthesis results from Synopsys Design Compiler with ASAP 7nm  library~\cite{asap}, the modulus only needs 2 cycles for calculation with 9-bit bandwidth, which is much quicker than memory access. Since the record protector only works upon the memory access of a \texttt{\small load}, the modulus calculation latency can be ignored through parallel calculation.

In summary, the hardware consumption is reasonable when \mydef{} is implemented in a modern 64-bit processor.

\section{Conclusion}
\label{sec:conclusion}


In this work, a secure prefetcher named \mydef{} is proposed, which can defeat cache side channel attacks while maintaining or even improving performance. In \mydef{}, Scale Tracker (ST), Access Tracker (AT), and Record Protector (RP) are designed to predict the eviction cachelines according to the victim's memory access during phase 2, predict the attacker's access patterns during phase 3, and increase the robustness, respectively.
The security is increased by prefetching the eviction cachelines that can confuse the attacker.
Experiments on Flush+Reload, Evict+Reload, and Prime+Probe prove the effectiveness and robustness of our defense. Besides, the average performance is also increased by the accurate prediction, according to the evaluations on SPEC CPU 2006 and 2017 benchmarks.

\ifCLASSOPTIONcaptionsoff
  \newpage
\fi



%


\bibliographystyle{IEEEtran}
\bibliography{ref}

\end{document}